\documentclass[journal abbreviation, manuscript]{copernicus}

\begin{document}
\nolinenumbers % for arxiv

\title{Maximum updraft velocity beyond CAPE: the role of boundary layer dynamics and pressure perturbations}

\Author[1][bety.pechacova@ista.ac.at]{Bety}{Pechacova}
\Author[1]{Alejandro}{Casallas}
\Author[2,3]{Tom}{Beucler}
\Author[4]{Lokahith}{Agasthya}
\Author[1]{Caroline}{Muller}

\affil[1]{Institute of Science and Technology Austria, Klosterneuburg, Austria}
\affil[2]{Faculty of Geosciences and Environment, University of Lausanne, Lausanne, Switzerland}
\affil[3]{Expertise Center for Climate Extremes, University of Lausanne, Lausanne, Switzerland}
\affil[4]{Department of Meteorology and Geophysics, University of Vienna, Vienna, Austria}

\runningtitle{Maximum updraft velocity beyond CAPE}

\runningauthor{Pechacova et al.}

\received{}
\pubdiscuss{} %% only important for two-stage journals
\revised{}
\accepted{}
\published{}

\firstpage{1}

\maketitle

\begin{abstract} 
Deep convective updraft velocities influence precipitation extremes, lightning, and the planetary energy budget. Convective Available Potential Energy (CAPE) is widely used to explain maximum updraft velocity ($w_{\max}$), but updrafts are also influenced by entrainment, boundary layer dynamics, pressure perturbations, and condensate loading. However, the relative importance of these processes and how they interact to set $w_{\max}$ in individual clouds remains unclear. Here, we use neural networks validated across climate regimes to identify generalizable pre-storm predictors and in-cloud diagnostics of $w_{\max}$ in individual tracked clouds. We then distill these into compact, physically interpretable equations, using idealized radiative–convective equilibrium simulations spanning a range of sea surface temperatures and radiative cooling rates. For pre-storm prediction, CAPE and local mean boundary layer vertical velocity ($\overline{w_{\mathrm{bl}}}$) together explain nearly half the variance in $w_{\max}$ across regimes ($R^2=0.47$). CAPE captures regime-mean differences, but it has little predictive value within a single simulation, where $\overline{w_{\mathrm{bl}}}$ captures cloud-to-cloud variability, including low $w_{\max}$ even at high $\mathrm{CAPE}$ values. At the time of peak intensity, a simple approximate Bernoulli-like invariant combining maximum pressure perturbation and maximum cloud condensate explains 89\% of the variance ($R^2=0.89$). The tight link between $w_{\max}$ and pressure perturbation is consistent with the sticky-thermal view and highlights dynamic pressure effects, often neglected in updraft theories. Both relationships remain skillful in additional simulations at horizontal grid spacings down to 125~m. These results highlight $\overline{w_{\mathrm{bl}}}$ as an important regulator of convective intensity alongside CAPE, and indicate that dynamic pressure plays an important role within individual updrafts.
\end{abstract}

\introduction  

Deep moist convection plays a central role in the Earth's climate system, strongly influencing the planetary energy budget \citep{bony_anvil_2016, ipcc_energybudget_2021} and driving extreme weather including intense rainfall, lightning \citep{romps_evaluating_2019}, and damaging winds \citep{muller_extremes_2020}. A key descriptor of deep convective intensity is the vertical velocity within updrafts. The strongest deep convective storms exhibit the largest updraft velocities, which are also closely linked to enhanced ice production and electrification \citep{singh_increases_2015, parodi_theory_2009}.

Historically, Convective Available Potential Energy (CAPE) has been widely used to predict thunderstorm intensity, updraft strength and vertical velocity \citep{Emanuel1994, johns_severe_1992, zipser_cumulonimbus_1980}. CAPE is a measure of the maximum potential energy available to be converted to kinetic energy in the adiabatic ascent of an undiluted moist parcel. Given that it is an energy, the updraft velocity is then assumed to scale with CAPE as $w \sim \sqrt{\mathrm{CAPE}}$. However, using CAPE alone to understand updraft dynamics suffers from a few drawbacks. CAPE provides only an upper limit to the acceleration that a moist parcel under adiabatic ascent undergoes, thus being an over-prediction for the vertical velocity of all but the most intense updrafts. The adiabatic idealization ignores irreversible processes, chiefly the dilution of moist updraft air parcels by turbulent mixing, a process known as entrainment. It is also known that observed temperature profiles of clouds do not exactly follow the temperature profile that a moist-adiabatic rising parcel of air would follow. Therefore, previous work has proposed modified versions of CAPE that incorporate entrainment rates, improving agreement with predicted and observed updraft velocities \citep{peters_analytic_2023, singh_influence_2013, singh_increases_2015, zhang_effects_2009, zhou_spectralplume_2019}. However, these approaches have not been extensively evaluated at the individual cloud level. Entrainment in deep convection is also more complex than continuous mixing with environmental air assumed in such plume models. Updrafts often consist of chains of successive thermals in which entrainment is localized, leaving parts of the updraft relatively undiluted \citep{morrison2020thermal, peters2020thermal}, and their cores can entrain humid air from a surrounding shell \citep{hannah2017entrainment} or cloudy air left by earlier thermals \citep{moser2017influence}, which dilutes them less than environmental air.

Within convective cores, vertical motion is tightly coupled to condensation, condensate loading, pressure adjustments, and various mixing and microphysical processes. Many of these are not captured by the potential buoyant acceleration assumed when calculating CAPE. Condensation drives vertical acceleration through latent heating, and vertical motion in turn drives condensation by lifting the air parcel. Once formed, hydrometeors also influence parcel buoyancy, and links between hydrometeor terminal fall speeds and vertical velocity have been proposed and examined \citep{parodi_theory_2009, singh_increases_2015}. Pressure perturbations also play multiple roles within updrafts, building up as a dynamical response to the flow and entering the vertical momentum balance through buoyancy and the pressure-gradient force. The importance of pressure effects in thermals and updrafts has been debated, with some studies neglecting them and others finding them to substantially regulate vertical motion \citep{jeevanjee_effective_2016, leger_simple_2019, sherwood_slippery_2013, hernandez-deckers_numerical_2016, romps_sticky_2015, kuo_anelastic_2025}.

In addition to thermodynamic potential (through CAPE) and in-cloud processes, convective initiation and other factors in the early development of an updraft can influence subsequent updraft intensity. Cold pools and boundary-layer heterogeneity can lead to localized convergence and lifting, providing mechanical and thermodynamic triggering at cloud base \citep{tompkins_coldpools_2001,torri_mechanisms_2015, fuglestvedt_coldpools_2020,abramian_extreme_2023, casallas_sensitivity_2025}. While these triggering processes mainly influence convective timing and organization, the convergence strength, lifting speed, or other boundary layer properties could also impact the vigor of updrafts \citep{agasthya_moist_2025}.

\begin{figure*}[t]
\includegraphics[width=0.9\textwidth]{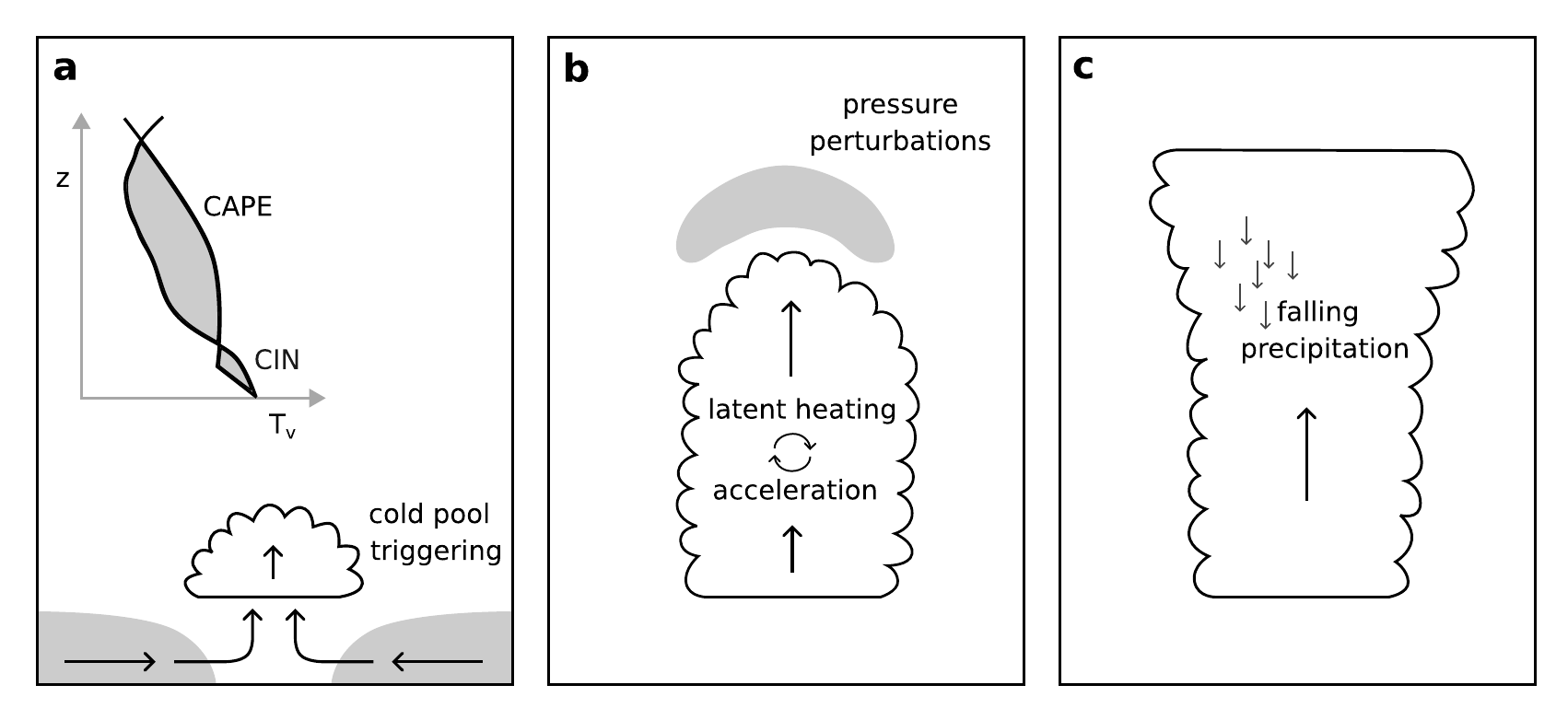}
\caption{Schematic overview of key processes influencing maximum updraft velocity in deep convection. Panels illustrate (a) thermodynamic control via CAPE and CIN and boundary-layer triggering, (b) latent heating and associated acceleration with pressure perturbation response, and (c) effects of condensate loading and hydrometeor terminal velocity. These processes may act simultaneously and interact within individual convective clouds.
}
\label{fig:mechanisms}
\end{figure*}

The processes discussed above and summarized schematically in Fig.~\ref{fig:mechanisms} are not mutually exclusive and may act simultaneously and interact within a single cloud. Although each of the mechanisms has theoretical or numerical support, they are often examined in isolation or evaluated using simulation-level statistics and metrics. As a result, their relative importance for extreme updraft velocities remains unclear, especially at the scale of individual clouds. It is not well established how much variability in updraft velocity can be explained by the pre-storm environment alone and to what extent in-cloud dynamical and microphysical structure provides additional information. Likewise, it is uncertain whether a small subset of physically interpretable variables can capture this variance or whether many variables are needed.

In this study, we develop a new approach (summarized in Fig.~\ref{fig:methods}), centered around data-driven equation learning, to improve our understanding of maximum updraft velocity, $w_{\max}$. We track clouds within radiative–convective equilibrium (RCE) simulations spanning a range of climate regimes, with different sea surface temperatures (SSTs) and radiative cooling rates. We first investigate a broad set of candidate lagged predictors describing the pre-storm and early-storm conditions. Second, we analyze a set of concomitant in-cloud diagnostics. In both cases, we use machine-learning (ML) based feature selection to find the most informative feature sets, either those associated with pre-storm conditions, or those concomitant with extreme updraft speeds. We employ cross-regime validation to ensure we select features generalizable across simulations and so likely to represent underlying physical mechanisms that are climate-invariant \citep{beucler2024climateinvariant}. Then, we use equation discovery on the selected features to distill the relationships learned by the ML models into compact, physically interpretable equations linking the dominant pre-storm predictors or concomitant in-cloud diagnostics to $w_{\max}$. Lastly, we analyze the selected feature sets and discovered equations to learn and form hypotheses about the important physical processes linked to the maximum updraft velocity.

Here we use RCE, which has long served as a fundamental framework for studying moist convection \citep[e.g.,][]{manabe_thermal_1964,held_rce_1993, wing_rcemip_2018,jeevanjee2022manabe}, including previous work on updraft velocity scaling \citep{singh_increases_2015, agasthya_moist_2025}. RCE is well suited to our goals for three reasons. First, in the absence of large-scale forcing, wind shear, and surface heterogeneity, variability in updraft velocities arises from a smaller set of fundamental processes, so relationships identified by equation discovery are more readily interpretable in physical terms. Second, RCE allows us to probe a wide range of climates in a controlled way: our simulations span five regimes with sea surface temperatures of 290--310~K and radiative cooling rates of 0.75--3~K\,day$^{-1}$, producing large variance in $w_{\max}$ and in candidate predictors and diagnostics. Such a range cannot be sampled with observations, and remains impractical with global storm-resolving models (GSRMs). It provides a stringent test of whether a discovered relationship is robust across climates rather than specific to one regime. Third, the idealized setting allows direct comparison with simple theoretical models of convective updrafts where these apply, connecting data-driven results to existing theory.

The remainder of the paper is organized as follows. Section~\ref{sec:methodology} describes the methodology, including the model configuration, cloud tracking, predictor definitions, and ML framework. Section~\ref{sec:resultsanddiscussion} discusses the results, first for the pre-storm predictive features and equations, then for the in-cloud diagnostic features and equations, and finally their sensitivity to horizontal resolution. We conclude with a summary and discussion of the implications of our findings in Sect.~\ref{sec:conclusion}.

\section{Methodology}\label{sec:methodology}

\begin{figure*}[t]
\includegraphics{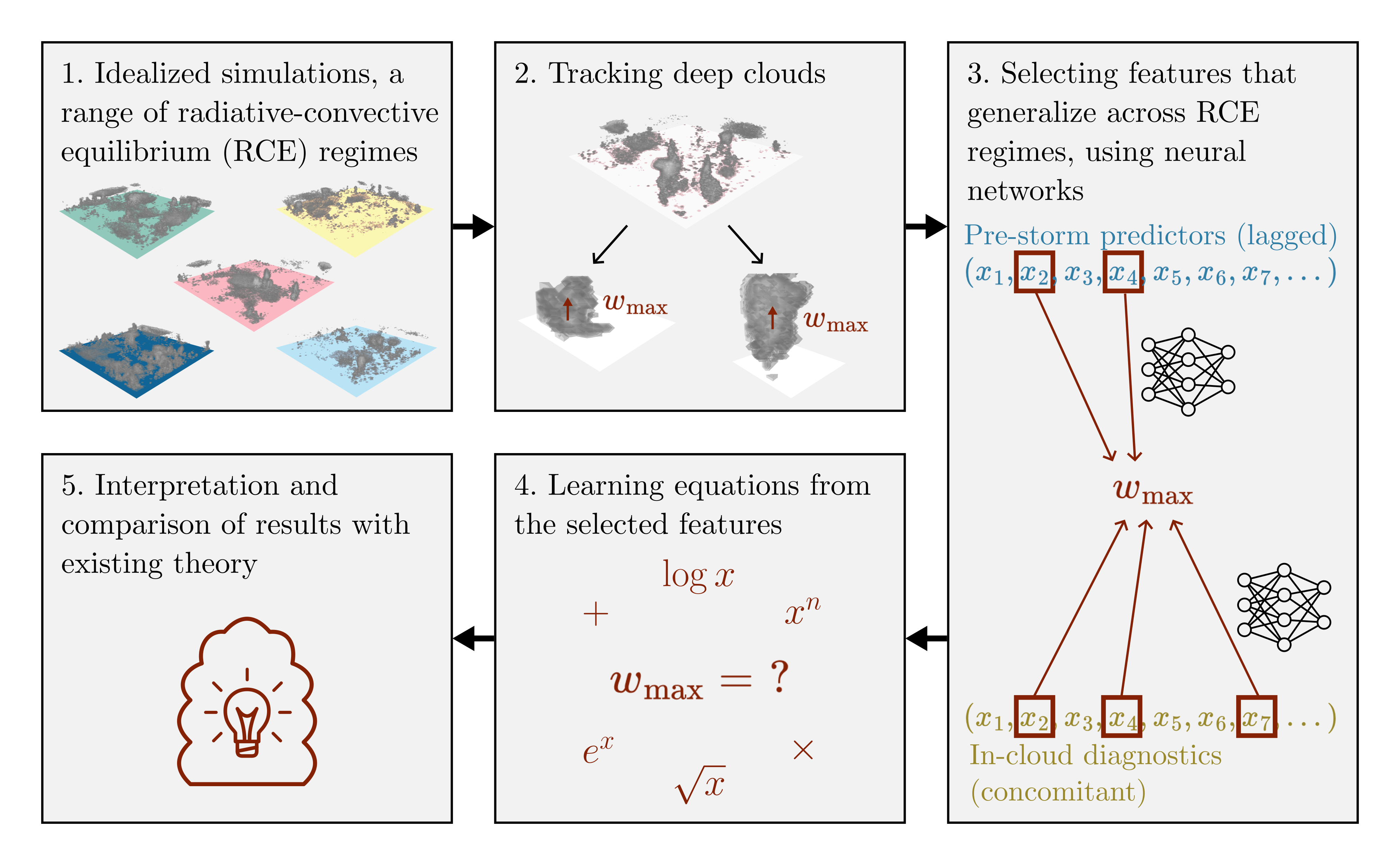}
\caption{Overview of the methodology framework. Starting from idealized radiative--convective equilibrium simulations (1.), individual deep convective clouds are tracked and their maximum updraft velocity $w_{\max}$ is identified (2.). Neural-network-based feature selection, with skill evaluated on simulations withheld from training, then identifies the pre-storm (lagged) predictors and in-cloud (concomitant) diagnostics that generalize across regimes (3.). Finally, equation discovery is applied to these selected features to obtain compact, physically interpretable relationships for $w_{\max}$ (4.), followed by interpretation in the context of known physical processes (5.).}
\label{fig:methods}
\end{figure*}

This section outlines the simulation design and the equation discovery methodology inspired by several other studies \citep{beucler_distilling_2024, grundner_data-driven_2024, hafner_rogue_2023, shamekh_rainarea_2026, zanna_ocean_2020}. In brief, our approach relies on identifying individual clouds in idealized cloud-resolving simulations and then implementing a data-driven equation discovery framework, concluding with a physical interpretation of the results. The steps are summarized in Fig.~\ref{fig:methods}.

\subsection{Model configuration}\label{sec:model_config}

We performed simulations in RCE using version 6.10.8 of the System for Atmospheric Modeling (SAM) \citep{khairoutdinov_cloud_2003}, with a set-up largely following \citet{agasthya_moist_2025}. Horizontally, the domain spans 128$\times$128~km with 1~km resolution and doubly periodic boundary conditions. The height of the domain is 27~km, with the upper third (above 18~km) consisting of a sponge layer with Newtonian damping to absorb gravity waves. There are 64 levels between the first atmospheric level (37.5~m) and 18~km, with vertical resolution finer near the surface and coarsening with altitude, from tens of meters in the planetary boundary layer to 400~m in the mid-troposphere; resolution coarsens further still within the sponge layer above. We use single-moment microphysics and a fixed SST. We output the 3D fields time-averaged over 5-minute intervals.

A constant radiative cooling rate $R$ is prescribed. At each timestep, the pressure level where the domain-mean temperature first drops below 200~K is diagnosed and defines the tropopause. The cooling is smoothly tapered to zero above this level using a hyperbolic tangent transition with a 50~hPa width. Temperatures below 200~K are relaxed toward 200~K with a 2-day Newtonian timescale.

\subsection{Experimental design}\label{sec:experiment_design}

We ran five experiments with different combinations of SST and $R$ values to represent different climate regimes:

\begin{center}
        \begin{tabular}{lcc}
        Simulation label & SST (K) & $R$ (K\,day$^{-1}$)\\
        CTRL & 300 & 1.50\\
        CTRL\_2$R$ & 300 & 3.00 \\
        CTRL\_0.5$R$ & 300 & 0.75 \\
        CTRL\_--10K & 290 & 1.50 \\
        CTRL\_+10K & 310 & 1.50 \\
    \end{tabular}
\end{center}

The simulations were run until they reached equilibrium, and we used enough steady-state data to yield around 2000 clouds per simulation (50--250 days, depending on the simulation).

\subsection{Cloud tracking}\label{sec:tracking}

We tracked clouds using a Lagrangian tracker adapted from \citet{casallas_airpollution_2024} and extended to four dimensions (time and three spatial dimensions). The tracker is applied to the 5-minute averaged 3D field of the non-precipitating cloud condensate mixing ratio (liquid and ice), with grid points exceeding $10^{-5}\mathrm{~kg~kg}^{-1}$ defined as cloudy. Individual objects are separated with the watershed algorithm implemented in scikit-image in Python \citep{scikit-image}, seeded at local maxima of the distance to the nearest non-cloudy point within a neighborhood of 20 output times, 20 vertical levels and 14$\times$14~km. Because the segmentation is performed in space and time simultaneously, each object is a cloud followed over its lifetime. Objects smaller than 40 space-time grid cells or touching the lateral domain boundaries are discarded.

The tracker returns all cloudy objects, most of which are not deep convective clouds but shallow or congestus clouds, anvil-only objects, or fragments separated from a neighboring cloud, which lack the initial growth phase. We select deep convective clouds in two steps. First, fixed thresholds remove objects that are clearly not of interest. We require a lifetime longer than 30~minutes, a vertical extent of more than 15 model levels over the lifetime, and, at the first output time of the object, a cloud base below 1.7~km and a cloud top below 5.3~km, so that the cloud is captured from its early growth.

These thresholds alone do not isolate deep convective clouds, and tightening them by hand did not give a satisfactory separation, partly because the geometrical properties that distinguish deep clouds vary between simulations (e.g., with tropopause height). In the second step, we therefore trained a classifier on manually labeled examples. We inspected 1463 objects passing the thresholds covering the five simulations, using time--height (Hovmöller) diagrams of vertical velocity averaged horizontally over the cloud. Objects showing coherent ascent from low levels to the upper troposphere were labeled as deep convective (344 objects, 24\%). We then trained an XGBoost classifier \citep{chen_xgboost_2016} on the simulation label and geometrical properties only: the minimum and maximum heights reached over the lifetime, height span, duration, maximum and time-mean volume, and the cloud-base and cloud-top heights at the first output time. Vertical velocity is not an input to the classifier. Hyperparameters were optimized using a grid search with stratified 5-fold cross-validation, selecting the model with the highest mean cross-validated F1 score (0.90). The final classifier was trained on all labeled objects and applied to all objects passing the thresholds.

\subsection{Definition of the maximum updraft velocity, $w_{\max}$}

For each tracked cloud, the maximum updraft velocity, $w_{\max}$, is defined as the largest grid-point value of vertical velocity attained within the cloud mask at any time during the cloud’s lifetime. Here, the cloud mask refers to all grid points identified as belonging to a given cloud by the tracking algorithm at each output time.

\subsection{Predictor definitions}

We perform two distinct analyses from this point on. The first uses lagged variables to identify predictors of $w_{\max}$, while the second uses diagnostics measured at the time of $w_{\max}$ to find quantities associated with the maximum updraft velocity, as described in more detail below. We note that the latter are not used for prediction and should not be interpreted as drivers of $w_{\max}$.

\subsubsection{Pre-storm predictors}\label{sec:prestorm_feature_defs}

A range of features related to the physical mechanisms discussed in the Introduction is defined to form the lagged predictor set, describing the pre-cloud and early growth phase. These include measures of the thermodynamic environment like CAPE (with and without entrainment, see below) and CIN, measures of moisture content like the relative humidity at different heights, and dynamic variables like the vertical velocity and horizontal winds. A full list of pre-storm features is included in Table~\ref{table:prestorm}. Spatially, these features are calculated over an 11$\times$11~km square around the location of $w_{\max}$ and they are lagged by 0.5 to 2 hours relative to the time of $w_{\max}$. Most of the resulting performance scores are not sensitive to the square size and lag choice, and the exceptions will be discussed in Sects.~\ref{sec:whatiswbl} and~\ref{sec:resolution_results}.

CAPE and CIN are computed from the vertically integrated buoyancy of an entraining plume. The plume is initialized with the temperature and humidity at the lowest model level (37.5~m), averaged over the 11$\times$11~km box, and integrated upward with a fractional entrainment rate $\epsilon(z) = \hat\epsilon/z$, where $z$ is the height above the surface. Entrainment is active above the lifting condensation level. The entrained air is assumed to have the box-mean properties. The calculation includes separate heat capacities for dry air, water vapor, liquid, and ice, a linear transition between liquid and ice between 233.15 and 273.15~K, and immediate fallout of all condensate. Buoyancy is computed from the difference between the plume temperature and the box-mean temperature, neglecting the virtual effect of water vapor. CAPE is then the integral of buoyancy from the level of free convection (LFC, defined as the lowest level at which the plume becomes positively buoyant) to the equilibrium level (EL, defined as the highest level at which it is positively buoyant). CIN is the integral of buoyancy below the LFC. We compute both for $\hat\epsilon = 0$, 0.5, and 1.0 (Table~\ref{table:prestorm}). For $\hat\epsilon = 0$, this calculation reduces to the standard pseudo-adiabatic CAPE, without the virtual correction. For simplicity, we use the label CAPE for all variants and discuss and specify the choice of $\hat\epsilon$ where relevant (see Sect.~\ref{sec:sqrtcape}).

\subsubsection{In-cloud diagnostics}\label{sec:incloud_feature_defs}

We also define a separate set of features measured at the time of $w_{\max}$ to characterize the structure of the cloud and surrounding environment at the time of peak intensity. These include the terminal velocity of hydrometeors, cloud condensate, temperature and pressure anomalies, buoyancy, and several environmental variables. Different statistics of these (mean, standard deviation, several percentiles) are taken either over the cloud or environment, where the cloud mask is defined by the tracking algorithm and the environment mask is defined as an 11$\times$11~km square around the location of $w_{\max}$, excluding any cloud regions. Table~\ref{table:incloud} shows the full list.

\subsection{Sequential feature selection and model evaluation}

We use sequential forward feature selection to identify generalizable subsets of predictors for $w_{\max}$, for the lagged and concomitant feature sets separately. Starting from an empty set, we iteratively add features. At each step, we add the feature from the remaining pool that maximally improves cross-validation performance and this procedure is repeated until all features are included.

The training procedure uses a Leave-One-Group-Out (LOGO) cross-validation scheme across the five simulations. In each fold, the model is trained on four simulations and evaluated on the fifth simulation. Because the coldest and warmest simulations are each held out in turn, this
procedure also tests extrapolation to conditions beyond the range seen in training. The models are fully connected feedforward neural networks (256–128–128–64 neurons) with ReLU activations. We used the adaptive gradient-based optimizer Adam \citep{kingma_adam_2014} to minimize mean squared error, and regularized via early stopping. Alternative architectures including larger networks and dropout layers were tested and produced comparable results. 

We quantify performance with the coefficient of determination, defined as:

\begin{equation}
    R^2 = 1 - \frac{\sum_i (y_i - \hat{y}_i)^2}
{\sum_i (y_i - \bar{y})^2},
\end{equation}
where $y_i$ denotes the true values, $\hat{y}_i$ the predictions, and $\bar{y}$ the mean of the true values over the evaluation set.

We considered two ways to compute $R^2$. For the \textit{across-simulations} score, predictions from all validation folds were concatenated and a single coefficient of determination was computed over the combined samples:
\begin{equation}
R^2_{\text{across}} 
= 1 - 
\frac{\sum_{s=1}^{N_s} \sum_{i \in s} (y_i - \hat{y}_i)^2}
{\sum_{s=1}^{N_s} \sum_{i \in s} (y_i - \bar{y}_{\text{across}})^2},
\end{equation}
where $\bar{y}_{\text{across}}$ denotes the mean over samples from all simulations, $s$ is a simulation label, and $N_s=5$ is the number of simulations. $R^2$ can also be computed separately for each held-out simulation and then averaged to get the \textit{within-simulations} score:
\begin{equation}
R^2_{\text{within}} 
= \frac{1}{N_s} \sum_{s=1}^{N_s}
\left[
1 - 
\frac{\sum_{i \in s} (y_i - \hat{y}_i)^2}
{\sum_{i \in s} (y_i - \bar{y}_s)^2}
\right],
\end{equation}
where $\bar{y}_s$ denotes the mean of the true values within simulation $s$.

The \textit{across-simulations} $R^2$ measures both the ability of a model to capture differences between different simulations and differences between different clouds within one simulation. In contrast, the \textit{within-simulations} $R^2$ only measures the average performance within a simulation, so the ability to capture cloud-to-cloud variability.

We use the \textit{across-simulations} $R^2$ as the main measure of skill unless otherwise specified and the number of features as a measure of model complexity. Following \citet{beucler_distilling_2024}, we place each model in the skill-complexity plane defined by these metrics. A model lies on the Pareto front if no other model achieves higher skill with equal or lower complexity. Thus, the front shows how much skill an additional feature adds and what the trade-offs are between model complexity and performance. We choose the size of the feature subsets for subsequent analysis by considering which features yield only marginal gains in skill and which features are consistently selected across repeated runs of the feature selection procedure.

\subsection{Equation learning}

The neural networks identify which features generalize across regimes and how much skill these features provide, but they are not interpretable. We thus use equation discovery to distill the relationships learned by the networks into explicit equations. The search is restricted to the feature subsets selected under cross-regime validation. This differs from applying symbolic regression in-sample to the full candidate pool, where the search could return relationships specific to the training simulations rather than relationships that hold across regimes.

We use two methods, each applied on scaled data with equal representation from each simulation. Pre-storm features are divided by their mean value, in-cloud features are standardized using z-scoring, and the target, $w_{\max}$, is not scaled.

First, we apply a symbolic regression algorithm using the PySR library \citep{cranmer_pysr_2023}. PySR is based on genetic programming and searches a space of functions defined by a set of operators (in our case $\{+,~-,~\times,~\div,~(\cdot)^{(\cdot)},~\exp,~\log\}$), variables, and parameters determining and limiting the complexity of possible expressions. The result is a set of Pareto-optimal equations balancing accuracy and complexity as defined by the PySR settings.

Second, we use the pySINDy library \citep{kaptanoglu_pysindy_2022} to apply a sequentially thresholded least squares algorithm on polynomial terms up to degree 5. This algorithm works by iteratively performing least squares regression with optional ridge regularization and masking out any weights below a given threshold \citep{brunton_discovering_2016}. We run the algorithm with a range of thresholds and regularization strengths to yield several equations with different numbers of terms.

\subsection{Equation baselines and evaluation}\label{sec:eq_evaluation}

To compare and evaluate the discovered equations, we use Pareto fronts with the number of features and the number of equation parameters as complexity metrics and $R^2$ as the performance metric. We include linear regression fits and baseline models, and use the L-BFGS-B algorithm \citep{zhu_lbfgsb_1997}, with multiple random restarts and keeping the lowest-error fit, to optimize any constants in all the equations, using the whole dataset. 

To select the best predictive pre-storm and diagnostic in-cloud equations from these Pareto planes, we also consider physical plausibility and constraints. Specifically, for the pre-storm case, we require that with all else fixed, the maximum vertical velocity does not decrease with CAPE, i.e., that $\partial w_{\max} / \partial \mathrm{CAPE} \geq 0$ (for $\mathrm{CAPE} > 0$), leaving two equations of different complexity (see Sects.~\ref{sec:sqrtcape} and \ref{sec:prestorm2vareq}). Among the in-cloud candidate equations, there were two close contenders, both with four parameters and a similar $R^2$, and from those we selected the one with a simpler analytical and also more physically justifiable form (see Sect.~\ref{sec:inclouddiscussion}).

To assess the robustness of the selected equations and their sensitivity to any particular simulation, the coefficients are additionally re-optimized in a LOGO cross-validation scheme, following the same split used for feature selection (in each fold, coefficients were fit on four simulations using L-BFGS-B and evaluated on the held-out fifth simulation). The resulting out-of-sample performance is reported in Appendix~\ref{appendix:eqrobustness}. Throughout the main text, we report coefficient uncertainties from a simulation-stratified bootstrap, which is described in Appendix~\ref{appendix:eqrobustness}.

\subsection{Resolution sensitivity experiments}\label{sec:resolution_methods}

Three simulations were run with higher horizontal resolution: 500, 250 and 125~m, for the CTRL set-up (see Sect.~\ref{sec:experiment_design}). We kept the domain size, vertical resolution, and other choices described in Sect.~\ref{sec:model_config} the same. The only changes were the horizontal resolution and correspondingly the simulation timestep. We used 35--50 days of statistically steady state data for each simulation to yield more than 2000 clouds per simulation. 

We treat the data from the higher resolution simulations in two different ways before we evaluate the equations in each. First, we coarsen the raw output by block-averaging all 3D fields horizontally to a 1~km grid. We then repeat the same pipeline (tracking and feature calculation) as for the main 1~km resolution simulations. Second, we use the native resolution output. To keep the computational cost feasible, the cloud condensate field in the 125~m and 250~m runs is averaged to 500~m for tracking (with parameters in Sect.~\ref{sec:tracking} scaled to keep the same physical size). To get the cloud mask on the native grids, only cloudy grid cells (non-precipitating cloud condensate mixing ratio exceeding $10^{-5}\mathrm{~kg~kg}^{-1}$) within the 500~m cloud label are used. All features and $w_{\max}$ are then computed on the native grid.

We refit the coefficients of the best discovered equations on both the coarsened and native grid datasets and report the results in Sect.~\ref{sec:resolution_results}. The pre-storm equation (Eq.~(\ref{eq:pysrtwovar})) contains the boundary-layer velocity averaged over a square region around the location of $w_{\max}$ (see Sects.~\ref{sec:prestorm_feature_defs}, \ref{sec:prestorm2vareq}, and \ref{sec:whatiswbl} for detail) and we recompute this for different box sizes from one grid cell up to around 25~km to see the effect of box size at different resolutions, and report results both for the default box and for the box size that maximizes $R^2$ in each simulation. Repeating the entire cross-regime validation and feature selection procedure would require running 12 additional RCE simulations (3 per regime), which is beyond the scope of this study.

\section{Results and discussion}\label{sec:resultsanddiscussion}

\subsection{Pre-storm conditioning: thermodynamic environment and boundary layer dynamics}

The pre-storm conditions seem to provide a mostly regime-level envelope for $w_{\max}$, capturing nearly half of the variance with the first two selected features, CAPE and the mean vertical velocity in the boundary layer ($\overline{w_{\mathrm{bl}}}$), 30~minutes before $w_{\max}$. Adding more features does not yield significant improvements (see Fig.~\ref{fig:prestorm_and_incloud_paretos}). With symbolic regression, we found equations with these variables which retain this predictive power, including the algorithm rediscovering the theoretical scaling of $w_{\max}$ with $\sqrt{\mathrm{CAPE}}$.

The following two sections (\ref{sec:sqrtcape}, \ref{sec:zmax}) will discuss CAPE, a widely used quantity and the most informative lagged predictor from our results. First, we will look at the theoretical scaling of $w_{\max}$ with $\sqrt{\mathrm{CAPE}}$, rediscovered by the ML methodology, and then we explore why CAPE provides limited predictive skill in our dataset. The following two sections (\ref{sec:prestorm2vareq}, \ref{sec:whatiswbl}) will analyze our best learned predictive equation for $w_{\max}$ using CAPE and $\overline{w_{\mathrm{bl}}}$, including a discussion of the functional dependence on the predictors and the physical interpretation. Lastly, before moving on to the in-cloud diagnostics results, we will discuss the possible sources of the remaining uncertainty in Sect.~\ref{sec:prestormremaininguncertainty}.

\subsubsection{Utility and limitations of CAPE}\label{sec:sqrtcape}

CAPE was found to be the most informative and the first selected lagged feature in our dataset, though alone it only captures around 25\% of the variance in the cross-regime validation scheme, and within one simulation, it has no predictive value ($R^2 \approx 0$). The amount of entrainment in the CAPE calculation has only a small effect on performance, with the across-simulations $R^2$ increasing slightly from 0.21 for $\hat\epsilon = 0$ to 0.25 for $\hat\epsilon = 1.0$ (Table~\ref{table:epsilon}). This is because the effect of entrainment is mainly a rescaling of CAPE by a similar factor in all our simulations, with a strong correlation between CAPE with different values of $\hat\epsilon$ across individual clouds. The relative differences between regimes, including the increase of CAPE with warming, are therefore nearly the same for all $\hat\epsilon$, and the rescaling is absorbed by the fitted coefficients of the equations below (Table~\ref{table:epsilon}) and by the neural networks. In all results shown, unless stated otherwise, we use CAPE with $\hat\epsilon = 1.0$, calculated 0.5~h before the time of $w_{\max}$, which gives the highest score. The main points discussed are not qualitatively sensitive to the choice of lag or $\hat\epsilon$.

\begin{figure*}[t]
\includegraphics{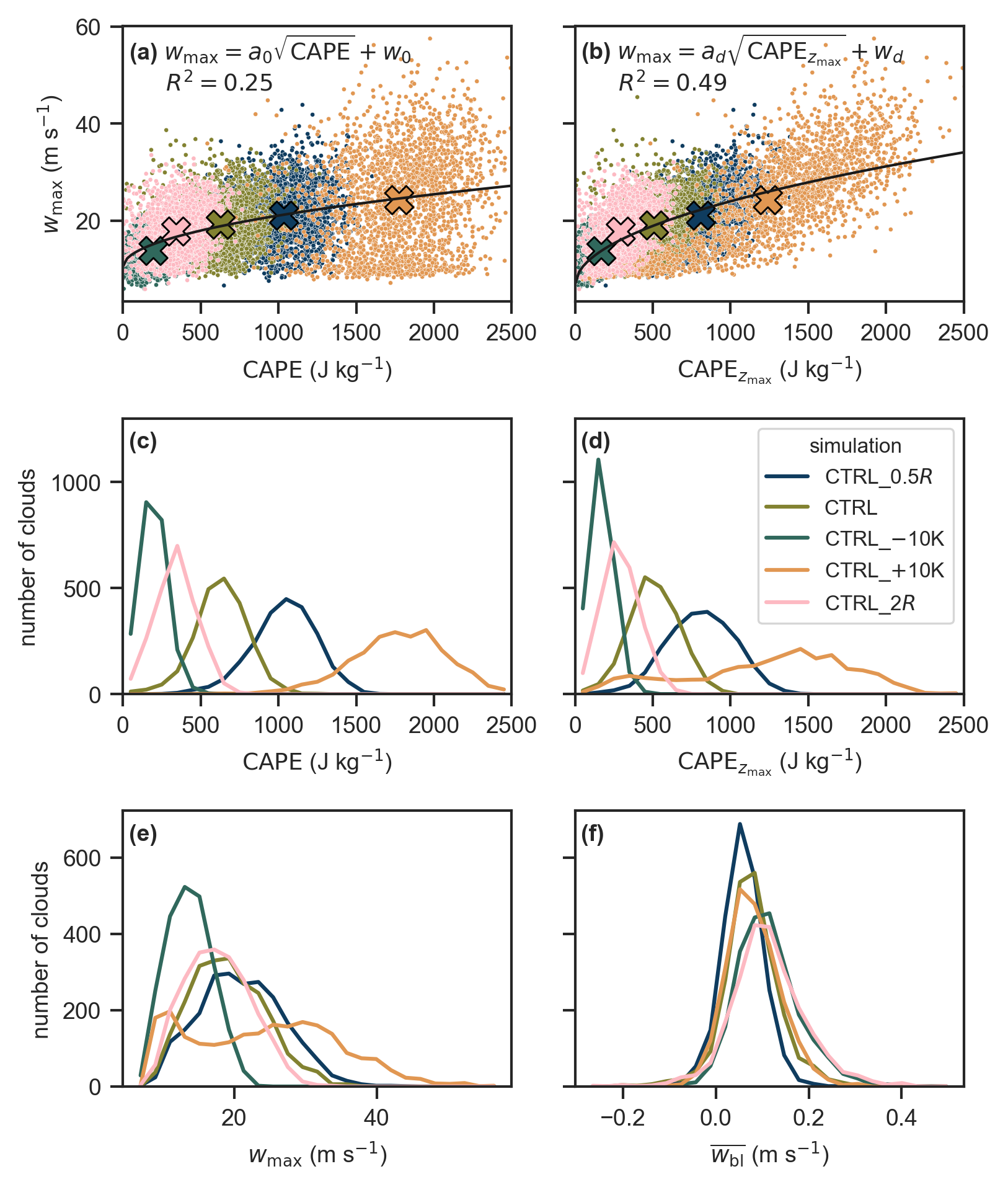}
\caption{
Relationship between $w_{\max}$ and $\mathrm{CAPE}$ ($\hat\epsilon=1.0$) across simulations. (a) Scatter of $w_{\max}$ versus $\mathrm{CAPE}$ with black line showing Eq.~(\ref{eq:capescaling}); crosses denote simulation means. (b) Same as (a) but with $\mathrm{CAPE}$ integrated up to $z_{\max}$, the height of $w_{\max}$, the black line showing predictions by the same equation with fitted coefficients $a_d=0.549 \pm 0.006,~w_d= 6.60 \pm 0.13 \mathrm{~m~s^{-1}}$. (c,~d) Histograms of $\mathrm{CAPE}$ and $\mathrm{CAPE}_{z_{\max}}$ for each simulation, in 25 uniform bins. (e,~f) Histograms of $w_{\max}$ and boundary-layer vertical velocity $\overline{w_{\mathrm{bl}}}$, in 25 uniform bins. Colors indicate different simulations.
}
\label{fig:capevscapez}
\end{figure*}

The PySR algorithm rediscovered the theoretical relationship of

\begin{equation}\label{eq:capescaling}
    w_{\max} = a_0 \sqrt{\mathrm{CAPE}} + w_0
\end{equation}

where $a_0 = 0.334 \pm 0.007$ and $w_0 = 10.44 \pm 0.15 \mathrm{~m~s^{-1}}$ are fitted constants, with CAPE in J~kg$^{-1}$. This result gives reassurance for the credibility of the methodology, but also shows CAPE alone is not sufficient in capturing the variability of $w_{\max}$ in our dataset. This equation describes the trend in simulation mean values well as shown by the crosses in Fig.~\ref{fig:capevscapez}a, and this is consistent with existing literature~\citep[e.g., ][]{singh_increases_2015}. This holds similarly for all three values of $\hat\epsilon$, with $R^2 \approx 0.94$ for predicting only the simulation means (Table~\ref{table:epsilon}). However, the variability within one simulation is much larger than the differences between simulation means as shown by the scatter of individual cloud points in the same plot and the heavily overlapping distributions of $w_{\max}$ in Fig.~\ref{fig:capevscapez}e. The distributions of CAPE (Fig.~\ref{fig:capevscapez}c), on the other hand, overlap much less, so something else must be influencing the maximum updraft velocity. 

Before discussing the best predictive equation, we will look more closely at some of the reasons why the utility of CAPE is limited in the next section.

\subsubsection{The role of the height where the maximum updraft velocity occurs}\label{sec:zmax}

When CAPE is diagnosed using the correct upper integration limit, defined by $z_{\max}$, the height of $w_{\max}$, the explained variance almost doubles relative to CAPE integrated up to the equilibrium level (both for Eq.~(\ref{eq:capescaling}) with re-optimized coefficients ($R^2=0.49,\ RMSE=5.1 \mathrm{~m~s^{-1}}$) and neural network models), as shown in Fig.~\ref{fig:capevscapez}b. This analysis is purely diagnostic, as $z_{\max}$ is not available as pre-storm information, but it highlights the potential importance of processes which set the height where the acceleration stops and so the maximum velocity occurs. 

However, the relative importance of $z_{\max}$ depends on the simulation set-up and could differ in other settings. CAPE is itself a function of many variables and so variations in the integration limits as well as properties of the initial parcel and the environment will contribute to variations in CAPE (and so $w_{\max}$). In our case of limited-domain RCE simulations with a fixed SST and $R$, the environment is relatively homogeneous and so the variability of $z_{\max}$ dominates over the thermodynamic spatiotemporal variations (see Appendix~\ref{appendix:capevariance}), limiting the utility of CAPE as a predictor of $w_{\max}$. 

These results show that even with a similar CAPE, $w_{\max}$ can vary widely, and this is partly due to variations in $z_{\max}$. Even where CAPE varies more widely than in our simulations, $z_{\max}$ variability is unlikely to be negligible. Thus, understanding or predicting $z_{\max}$ could improve pre-storm predictability.

\subsubsection{A predictive equation for the maximum updraft velocity with CAPE and $\overline{w_{\mathrm{bl}}}$}\label{sec:prestorm2vareq}

\begin{figure*}[t]
    \includegraphics{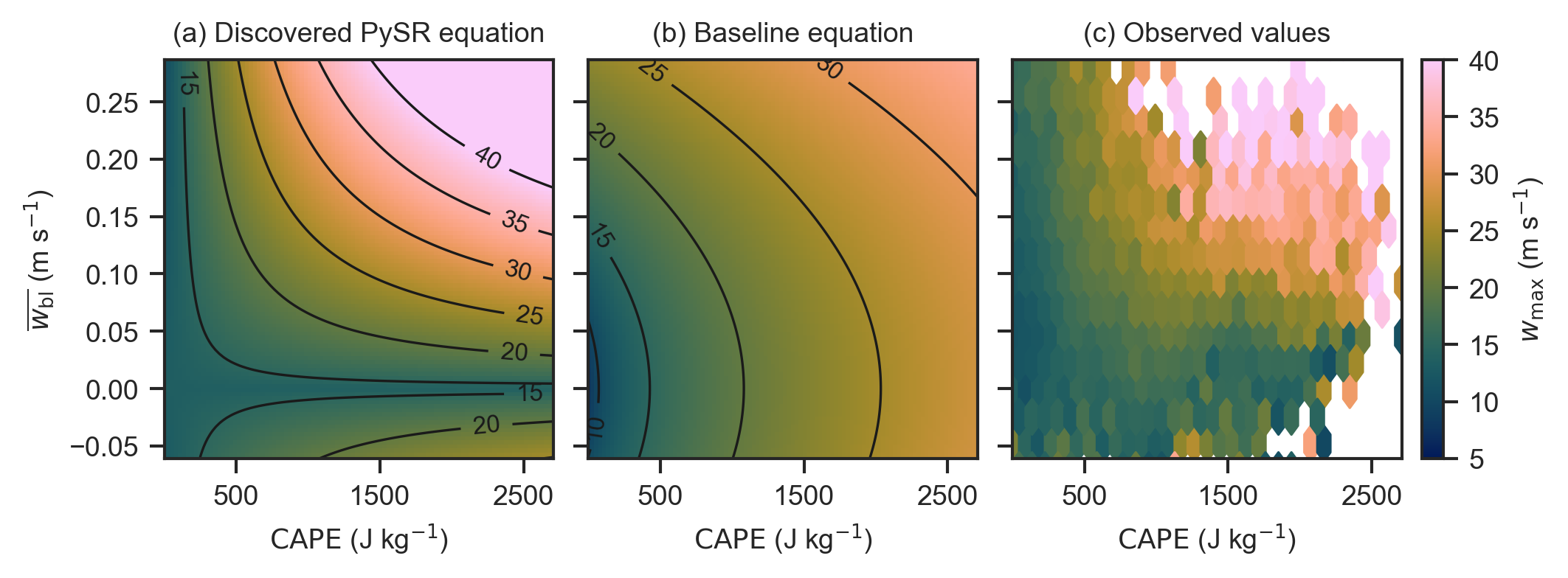}
    \caption{
    Comparison of predictive relationships for $w_{\max}$ as a function of $\mathrm{CAPE}$ and boundary-layer vertical velocity $\overline{w_{\mathrm{bl}}}$. (a) Discovered symbolic regression equation (Eq.~(\ref{eq:pysrtwovar})). (b) Baseline equation (Eq.~(\ref{eq:capetrigger})). (c) Observed $w_{\max}$ values from all simulations combined. Contours and color shading indicate $w_{\max}$ magnitude (m s$^{-1}$).
    }
    \label{fig:prestormeqs}
\end{figure*}

We can formulate a baseline equation by adding $\overline{w_\mathrm{bl}}^2$ as an initial kinetic energy into Eq.~(\ref{eq:capescaling}):

\begin{equation}\label{eq:capetrigger}
    w_{\max} = a_1\sqrt{\mathrm{CAPE} + b~ \overline{w_\mathrm{bl}}^2} + w_1
\end{equation}

where $a_1=0.409 \pm 0.006,~ b=(2.07 \pm 0.08) \times 10^{4},~ w_1=6.55 \pm 0.16 \mathrm{~m~s}^{-1}$ when fitted to the data. Eq.~(\ref{eq:capetrigger}) yields $R^2 \approx 0.32$ and $RMSE=5.9 \mathrm{~m~s}^{-1}$, a small improvement considering a second variable was added.

The best predictive equation for $w_{\max}$ learned by the data-driven framework, which performs much better at $R^2=0.47$ and $RMSE = 5.2\mathrm{~m~s}^{-1}$, is

\begin{equation}\label{eq:pysrtwovar}
    w_{\max} = w_2 \left[\, 1 + \left( \frac{|\overline{w_{\mathrm{bl}}}|}{w_\mathrm{ref}} \right)^{\!c} \log\!\left( \frac{\mathrm{CAPE} + \mathrm{CAPE}_0}{\langle \mathrm{CAPE} \rangle} \right) \right]
\end{equation}

where $w_2 = 13.53 \pm 0.14 ~\mathrm{m\,s^{-1}}$, $w_\mathrm{ref} = 0.120 \pm 0.003~\mathrm{m\,s^{-1}}$, $c = 0.778 \pm 0.023$, $\mathrm{CAPE}_0 = 710 \pm 10~ \mathrm{J\,kg}^{-1}$ are fitted constants and $\langle \mathrm{CAPE} \rangle = 796~\mathrm{J\,kg^{-1}}$ is the sample-mean CAPE used to normalize the input. CAPE and $\overline{w_{\mathrm{bl}}}$ are in J~kg$^{-1}$ and m~s$^{-1}$, respectively. 

The main difference between Equations (\ref{eq:capetrigger}) and (\ref{eq:pysrtwovar}) is the dependence of $w_{\max}$ on $\overline{w_{\mathrm{bl}}}$ and how that changes with CAPE, as seen in panels a and b in Fig.~\ref{fig:prestormeqs}. While the baseline equation describes $w_{\max}$ as mostly a function of CAPE with a correction from $\overline{w_{\mathrm{bl}}}$ with decreasing significance at higher CAPE values, the discovered equation shows a much stronger dependence on the boundary layer velocity, with small values of $w_{\max}$ even with high CAPE when $\overline{w_{\mathrm{bl}}}$ is small, consistent with the data (Fig.~\ref{fig:prestormeqs}c). Including $\overline{w_{\mathrm{bl}}}$ in the model in this way appropriately leads to the distributions of predicted $w_{\max}$ values in different simulations overlapping more heavily and the performance almost doubling compared to the CAPE-only scaling discussed in Sect.~\ref{sec:sqrtcape}. 

Clearly, boundary layer dynamics play a larger role in controlling $w_\mathrm{max}$ than a simple additive picture would suggest, with a strongly nonlinear interaction between CAPE and $\overline{w_\mathrm{bl}}$. In Eq.~(\ref{eq:pysrtwovar}), the dependence of $w_\mathrm{max}$ on CAPE is multiplied by a factor that increases with $|\overline{w_\mathrm{bl}}|$. Rather than only setting the initial kinetic energy (i.e., the triggering velocity at the top of the boundary layer, to which acceleration from CAPE is then added), $\overline{w_\mathrm{bl}}$ thus modulates how much of the energy available from CAPE is realized as maximum updraft velocity, so that for weak boundary-layer ascent $w_\mathrm{max}$ remains close to $w_2$ regardless of CAPE (Fig.~\ref{fig:prestormeqs}a). As a note, the presence of the natural logarithm, maybe physically surprising, should not be over-interpreted --- replacing it with a square root and refitting the equation yields $R^2=0.44$ and $RMSE = 5.3\mathrm{~m~s}^{-1}$, so while the nonlinear interaction is significant, the dependence on CAPE can be captured similarly well in different ways. What $\overline{w_\mathrm{bl}}$ physically represents is less clear, which we examine next.

\subsubsection{What is $\overline{w_\mathrm{bl}}$?}\label{sec:whatiswbl}

Unlike CAPE, the mean vertical velocity in the boundary layer alone does not generalize well from regime to regime, but it does have some predictive skill within one simulation ($R^2=0.31$ on average for a network trained and evaluated on samples from the same simulation). This is not surprising given the $\overline{w_{\mathrm{bl}}}$ distributions (Fig.~\ref{fig:capevscapez}f) are much more similar between the simulations than for $w_{\max}$ (Fig.~\ref{fig:capevscapez}e) and CAPE (Fig.~\ref{fig:capevscapez}c). Also unlike CAPE, its utility is lag-sensitive. Taking $\overline{w_{\mathrm{bl}}}$ 45~minutes prior instead of 30~minutes reduces the cross-validation $R^2$ of the CAPE and $\overline{w_{\mathrm{bl}}}$ neural network from 0.45 to 0.34.

Physically, $\overline{w_\mathrm{bl}}$ could be a measure of several factors. As it is an average over an $11 \times 11 \mathrm{~km}$ square in the boundary layer, it could be indicative of convergence and lifting related to cold pools \citep{fuglestvedt_coldpools_2020}, or the net upward mass flux, possibly dominated by the updraft size or speed. The former interpretation is consistent with \citet{Robinson2025}, who found that CAPE and near-surface convergence act jointly, through their product, although in their simulations this controlled the frequency rather than the intensity of deep convection. Conditioning the average on updraft ($w>0$) or cloud ($q_c>0$) grid points, or using the updraft or cloud volume fraction in place of $\overline{w_\mathrm{bl}}$, all result in neural network scores better than using CAPE alone, but significantly worse than CAPE with $\overline{w_\mathrm{bl}}$. Hence, $\overline{w_\mathrm{bl}}$ does not reduce to any single one of these cloud or updraft properties. 

One possible interpretation follows from the multiplicative structure of Eq.~(\ref{eq:pysrtwovar}), which resembles, though only structurally, the analytic expressions of \citet{Morrison2016}. There, the vertical velocity scales with $\sqrt{\mathrm{CAPE}}$ reduced by a factor depending on the updraft aspect ratio. This suggests that $\overline{w_{\mathrm{bl}}}$ might carry information on the updraft geometry at the time of $w_{\max}$, which is difficult to diagnose before the updraft is fully developed. Observations in the trades show that clouds rooted in merged, wider boundary-layer thermals are themselves wider and deeper \citep{bonyRoleThermals2025}, suggesting that boundary-layer structure can influence the width of developing clouds. A larger $\overline{w_{\mathrm{bl}}}$ could reflect such wider or merged thermals. Wider updrafts are less diluted by entrainment and convert CAPE into kinetic energy more efficiently \citep{peters_analytic_2023}, providing a possible route from $\overline{w_{\mathrm{bl}}}$ to $w_{\max}$. Testing this hypothesis would require diagnosing the updraft width directly.

Alternatively, $\overline{w_{\mathrm{bl}}}$ could act as a proxy for warm or moist anomalies in the air feeding the updraft, which are smoothed out in our CAPE because its initial parcel is averaged over the 11$\times$11~km box and the 5-minute output interval. A plume initialized with such an anomaly would have larger buoyancy throughout its ascent, increasing CAPE nonlinearly, which could contribute to the nonlinear interaction between CAPE and $\overline{w_{\mathrm{bl}}}$ found here.

A more detailed examination of the boundary layer is beyond the scope of this study but could clarify its role in more detail. Additionally, thanks to its straightforward definition, the predictive value of $\overline{w_\mathrm{bl}}$ could also be tested in other models or datasets.

\subsubsection{Sources of remaining uncertainty}\label{sec:prestormremaininguncertainty}

To summarize our pre-storm results, the best models explain nearly half of the variance in $w_{\max}$ across RCE regimes and use CAPE and the mean vertical velocity in the boundary layer, $\overline{w_\mathrm{bl}}$. CAPE is a widely and historically used predictor of updraft speeds, but in our set-up, it is not sufficient to predict the value of $w_{\max}$. CAPE is quite homogeneous within each of our simulations and very different in the different regimes, yet the distributions of $w_{\max}$ overlap heavily. So, the mean CAPE in a simulation corresponds well to the mean $w_{\max}$ but at an individual cloud level, there is a lot of remaining variability. Some of this is explained by $\overline{w_{\mathrm{bl}}}$, some by diagnosing the height of $w_{\max}$, and some of it is unexplained by any of the tested predictors. It could come from a combination of inherent stochasticity in the system, uncertainties from cloud and feature definitions, and any additional unaccounted-for predictors. Boundary layer dynamics have a strong influence on $w_{\max}$, limiting its values for many clouds in high CAPE environments and explaining some of the cloud-to-cloud variability within a regime. Nevertheless, what the exact role is remains an open question. 

\subsection{In-cloud diagnostics: Latent heating and pressure perturbation}\label{sec:inclouddiscussion}

\begin{figure*}[t]
\includegraphics{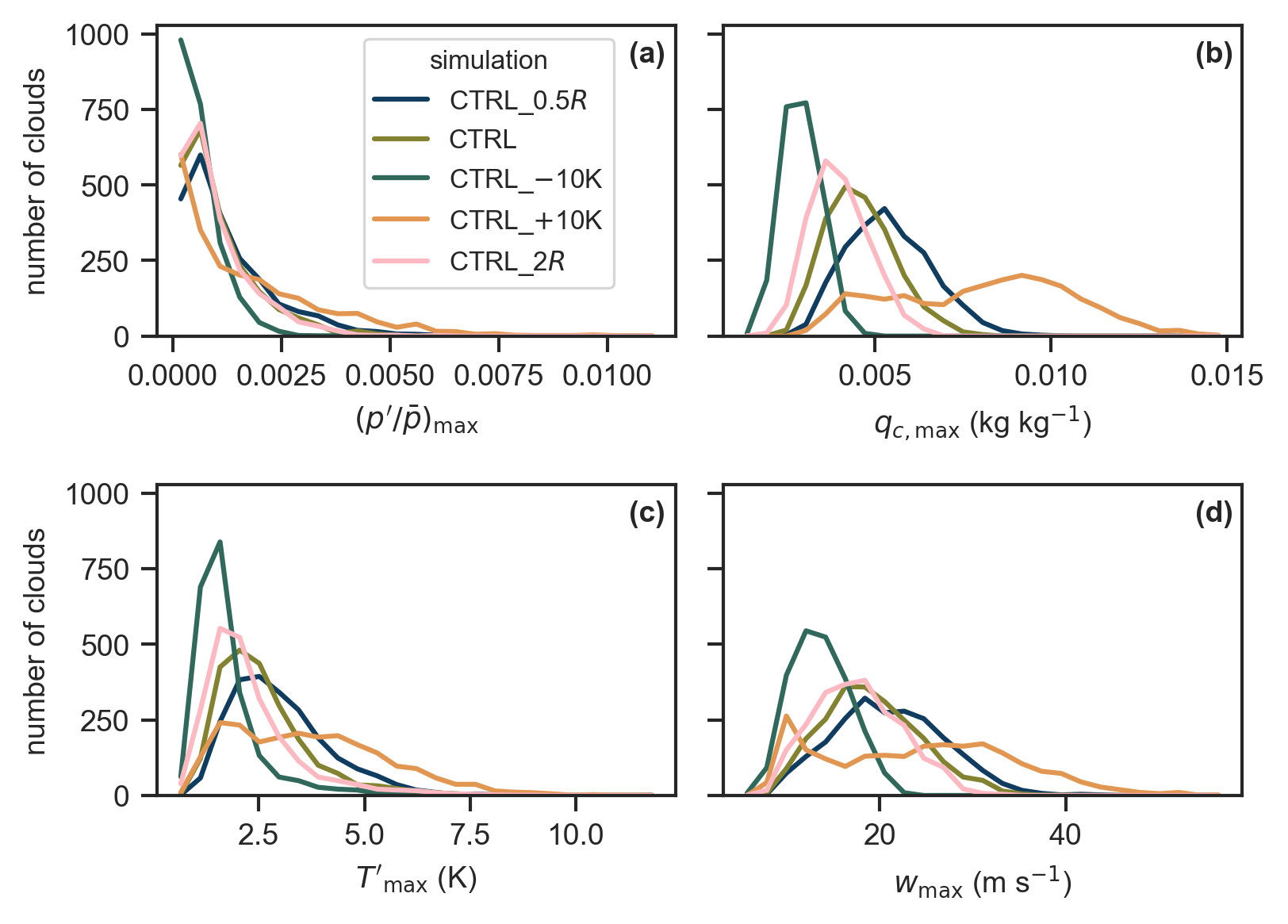}
\caption{
Histograms of selected in-cloud diagnostic features and $w_{\max}$ across simulations. Shown are (a) maximum pressure perturbation $(p'/\bar{p})_{\max}$, (b) maximum cloud condensate mixing ratio $q_{c,~\max}$, (c) maximum temperature perturbation $T'_{\max}$, and (d) maximum updraft velocity $w_{\max}$. Colors denote different simulations and histograms are calculated using 25 uniform bins for each variable.
}
    \label{fig:incloud_pdfs}
\end{figure*}

From the concomitant features, the top three selected by the ML procedure are: the maxima of the fractional pressure perturbation ($(p'/\overline{p})_{\max}$), cloud condensate mixing ratio ($q_{c,\max}$), and temperature perturbation ($T'_{\max}$) in the cloud, together capturing 91\% of the variance in the cross-validation scheme (see the full Pareto front in Fig.~\ref{fig:prestorm_and_incloud_paretos}, and the feature distributions in Fig.~\ref{fig:incloud_pdfs}).

We find that the feature selection consistently favors extrema-based diagnostics for $w_{\max}$, whether it is performed with cross-validation across simulations as described here, only with data from one simulation, or using data from all simulations for training and validation. The selected features are also the same when we introduce a small lag (5--10 minutes). The full in-cloud feature set (see Table~\ref{table:incloud}) also included distributional statistics such as means and higher moments so the dominance of maxima was not imposed a priori. The maximum updraft velocity thus scales robustly with other in-cloud extrema and as will be discussed later, these maxima are found in different locations in the cloud, making the association a description of more than just a localized relationship.

\begin{figure*}[t]
\includegraphics{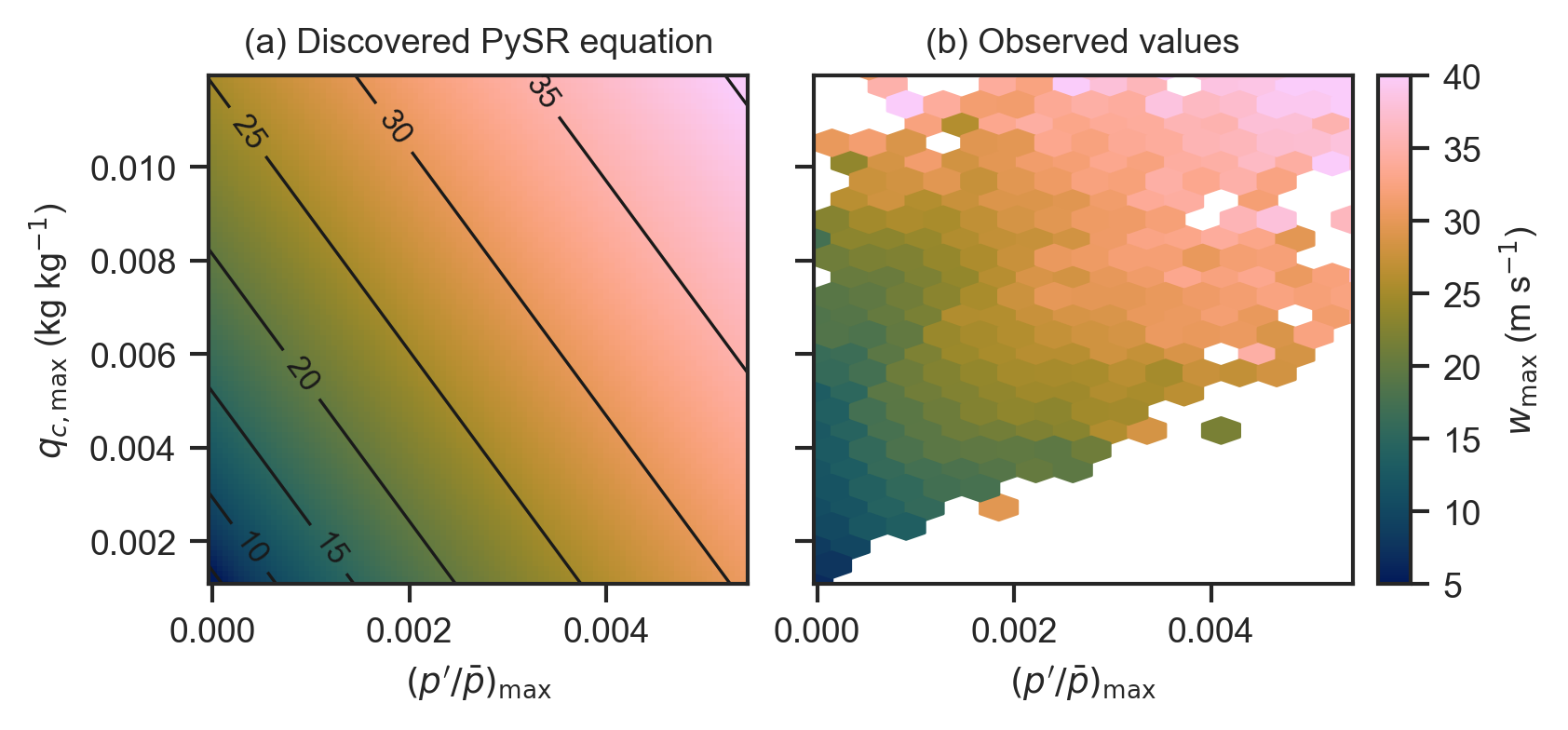}
    \caption{Relationship between $w_{\max}$ and key in-cloud diagnostics. (a) Predicted $w_{\max}$ from the discovered equation (Eq.~(\ref{eq:incloudpysr})) as a function of $(p'/\bar{p})_{\max}$ and $q_{c,\max}$. (b) Observed values from all simulations combined. Colors indicate $w_{\max}$ (m s$^{-1}$).
}
    \label{fig:incloudeqplot}
\end{figure*}

Applying equation discovery to the selected features, we find several simple equations that retain most of the skill of the neural networks. The best equation, with $R^2=0.89$ and $RMSE = 2.4 \mathrm{~m~s}^{-1}$ is:

\begin{equation}\label{eq:incloudpysr}
w_{\max}
=
\sqrt{\alpha \left(\frac{p'}{\bar p}\right)_{\max}
+
\beta q_{c,\max} + \gamma} + \delta
\end{equation}

with $\alpha= (2.07 \pm 0.04)\times 10^5 \mathrm{~m^2 ~ s^{-2}},\ \beta= 72.5 \pm 2.3 \mathrm{~m^2~s^{-2}(g~kg^{-1})^{-1}},\ \gamma= -21 \pm 10 \mathrm{~m^2~s^{-2}},\ \delta= -3.8 \pm 0.5\mathrm{~m~s^{-1}}$, and $q_{c,\max}$ in $\mathrm{g~kg^{-1}}$. 
The behavior of this equation in the $(p'/\overline{p})_{\max}$--$q_{c,\max}$ plane along with the true samples is shown in Fig.~\ref{fig:incloudeqplot}. The maximum updraft velocity increases with both $(p'/\overline{p})_{\max}$ and $q_{c,\max}$, and the magnitude of the two terms in the range of the dataset is comparable. We can rewrite Eq.~(\ref{eq:incloudpysr}) in a Bernoulli-like approximate invariant:

\begin{equation}\label{eq:incloudinvariant}
    I = 
    \frac{1}{2}\left(w_{\max} - w_{\mathrm{off}}\right)^2
   - \varepsilon_p \left(\frac{p'}{\bar p}\right)_{\max}
   - \eta_L L_v q_{c,\max}
   + \varepsilon_0
   \approx 0
\end{equation}

where $L_v$ is the latent heat of condensation, and the fitted coefficients become a velocity offset $w_\mathrm{off} = \delta$, energy scales $\varepsilon_p=\alpha/2$ and $\varepsilon_0=-\gamma/2$, and an empirical efficiency $\eta_L=\beta(2L_v)^{-1}$. While Eq.~(\ref{eq:incloudinvariant}) resembles the Bernoulli potential, it is not straightforward to interpret it as such; instead, it can be thought of as an approximate invariant in the sense that $I \approx 0$ at the time of peak intensity across clouds and regimes. Its form suggests that across climate regimes, diverse cloud states collapse onto a low-dimensional energy-like manifold involving peak velocity and the maxima of pressure perturbation and cloud condensate.

We interpret the discovered equation as capturing a scaling of the maximum updraft velocity with thermodynamic forcing from latent heating and with the dynamical pressure response to the flow. The following sections discuss these two processes, first the condensation and latent heating in Sect.~\ref{sec:latentheating} and then the pressure response in Sect.~\ref{sec:pp}. We conclude by bringing those together in Sect.~\ref{sec:incloudeqstructure} to rationalize the learned diagnostic relationship given by Eq.~(\ref{eq:incloudinvariant}) in the context of the other analyses and interpretation.

\subsubsection{Latent heating}\label{sec:latentheating}

The maxima of the cloud condensate and temperature perturbation ($q_{c,\max}$ and $T'_{\max}$) are likely proxies for the net effect of latent heating due to condensation (recall that although $T'_{\max}$ does not appear in Eq.~(\ref{eq:incloudinvariant}), it is among the top three selected features as discussed above). These moist processes are tightly coupled to vertical acceleration and motion within the updrafts (though our diagnostic analysis cannot determine the causal direction). 

\begin{figure}[t]
    \includegraphics[width=8cm]{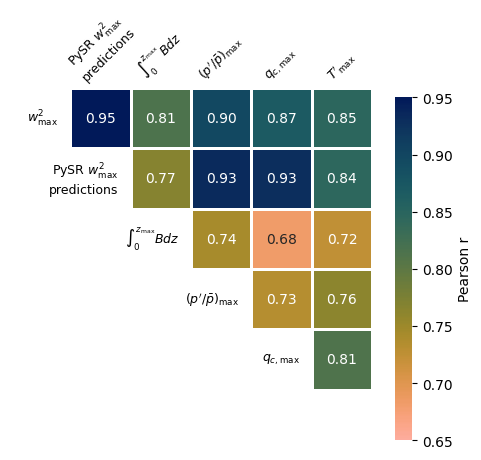}
    \caption{
    Correlation matrix between $w_{\max}^2$, predicted $w_{\max}^2$ from the discovered equation (Eq.~(\ref{eq:incloudpysr})), buoyancy integral $\int_0^{z_{\max}} B \,\mathrm{d}z$, and selected in-cloud diagnostic features. Colors indicate Pearson correlation coefficients, using samples from all simulations combined.
}
    \label{fig:corrmatrix}
\end{figure}

A more exact measure of acceleration from condensation would be a vertical buoyancy ($B$) integral, which has been used to diagnose vertical velocities in other studies \citep[e.g.,][]{agasthya_scaling_2024} and would yield the scaling $w_{\max}^2\sim \int_0^{z_{\max}} B \,\mathrm{d}z$. Here we diagnose the buoyancy profile at the time and location of $w_{\max}$, using the SAM definition of $B \approx g(T'/\overline{T}+0.608~q_v' - q_c-p'/\overline{p})$ \citep{khairoutdinov_cloud_2003}. Equation~(\ref{eq:incloudpysr}) is much simpler than such a computation and outperforms diagnosing $w_{\max}$ from a vertical integral of the buoyancy as well as the diagnostic $\mathrm{CAPE}_{z_{\max}}$ relation. Even any one of the three selected features is more strongly associated with $w_{\max}$ than $\int_0^{z_{\max}} B \,\mathrm{d}z$, despite those only being one statistical moment over the cloud (see Fig.~\ref{fig:corrmatrix}).

While CAPE is an idealized prediction of latent heating based on a moist adiabatic ascent, $q_{c,\max}$ and $T'_{\max}$ capture the realized latent heating taking into account the net effects of entrainment and mixing, redistribution of thermodynamic perturbations through the cloud vertically, and dynamical constraints including pressure-gradient and other forces. The fact that $q_{c,\max}$ and $T'_{\max}$ outperform CAPE and buoyancy-based predictors could suggest that the cumulative effect of these non-adiabatic processes is large enough to substantially modify the realized $w_{\max}$ from its undiluted-parcel value. These processes include the entrainment of humid or cloudy air from around the updraft core discussed in the Introduction, which occurs on the scale of individual thermals, with radii of a few hundred meters \citep{hernandez-deckers_numerical_2016}, and is therefore mostly subgrid at 1~km. However, Eq.~(\ref{eq:incloudpysr}) also holds at finer resolutions, where this mixing is better resolved (see Sect.~\ref{sec:resolution_results}).

The coupling between vertical motion and condensation is, however, not one-directional. Stronger updrafts also drive faster condensation \citep{grant_linear_2022}, and when condensate is produced faster than it is converted to precipitation, $q_{c,\max}$ also reflects the condensate accumulated due to fast ascent. Condensation and vertical motion are therefore best viewed as coupled processes, and the advantage of $q_{c,\max}$ over buoyancy-based predictors may partly reflect the effect of the updraft on the condensate rather than only the effect of latent heating on the updraft.

\begin{figure*}[t]
\includegraphics{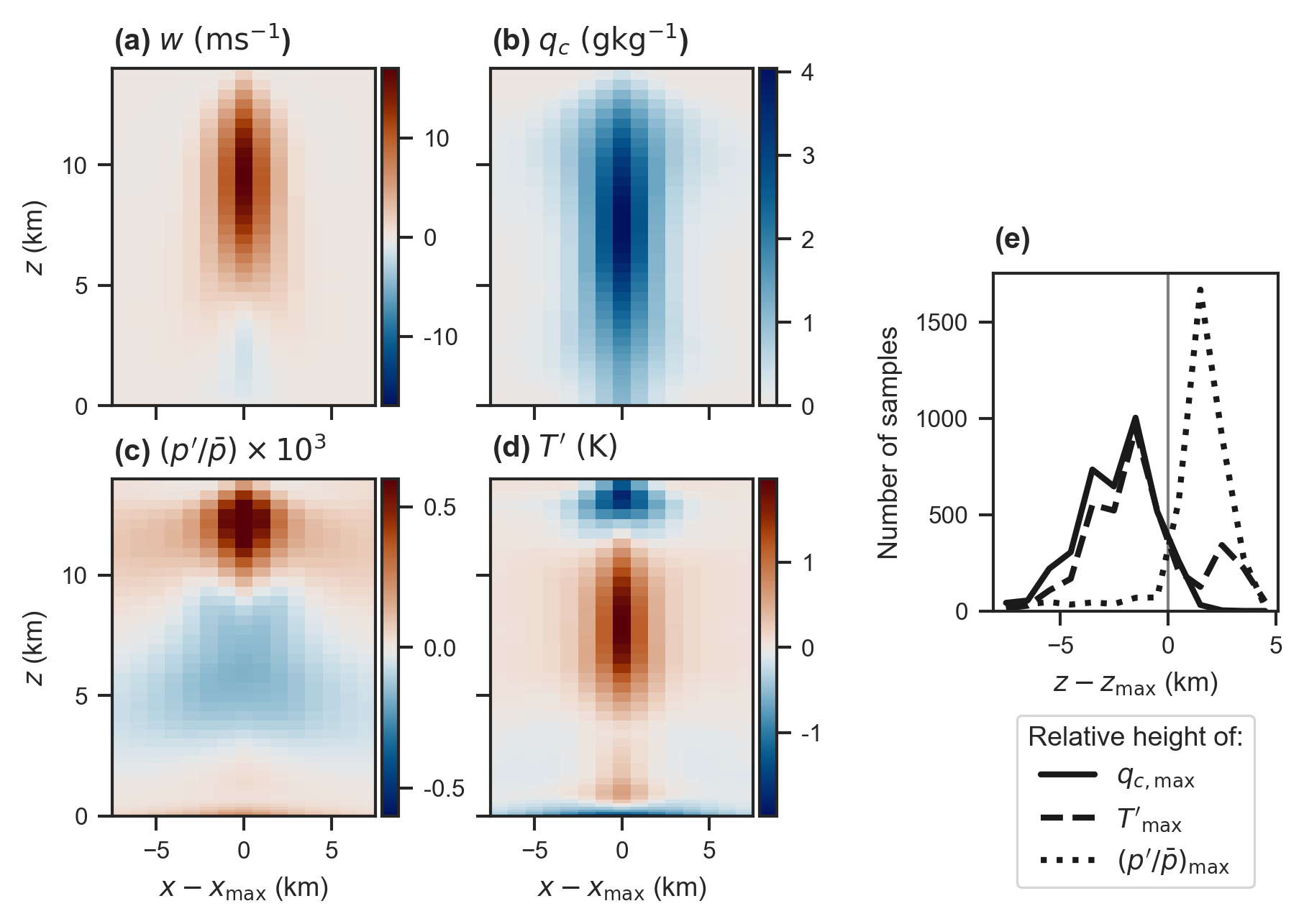}
\caption{
Composite structure of convective clouds at peak intensity in the CTRL simulation. (a--d) Horizontally centered composites at the location of $w_{\max}$ showing (a) vertical velocity $w$, (b) cloud condensate $q_c$, (c) fractional pressure perturbation $p'/\bar{p}$, and (d) temperature perturbation $T'$. (e) Histograms of vertical offsets between the height of $w_{\max}$ and the heights of the maxima of $q_c$, $T'$, and $p'/\bar{p}$ (in the CTRL simulation).
}
    \label{fig:composite}
\end{figure*}

$q_{c,\max}$ and $T'_{\max}$ are found below $w_{\max}$ in most clouds as shown in Fig.~\ref{fig:composite}e, and the fields of $q_{c}$ and $T'$ averaged over clouds mirror the shape of the updraft (compare panels a, b, and d in Fig.~\ref{fig:composite}). This is consistent with these variables indicating latent heating within the core of the cloud associated with vertical acceleration and any redistribution processes. Despite occurring at different heights from $w_{\max}$, these quantities scale closely with it, suggesting that the relationship given by Eq.~(\ref{eq:incloudinvariant}) reflects properties of the cloud as a whole, rather than a local balance at a single level or within a limited region.

Lastly, the functional relationship between $q_{c,\max}$ and $w_{\max}$ in Eq.~(\ref{eq:incloudinvariant}) is also consistent with the cloud condensate being a proxy for the net condensation and heating, and hence buoyancy and acceleration, in the cloud. Buoyancy in the updraft is dominated by the temperature anomaly term, $gT'/\overline{T}$, and the temperature excess is expected to scale as the condensate, $c_pT' \approx L_v q_c$ if they are both primarily driven by condensation processes. Hence, assuming $\overline{T}$ varies slowly compared to $q_c$ with height and the column-integrated cloud condensate roughly scales as $q_{c,\max}$, we recover $w_{\max}^2 \sim q_{c,\max}$ as in the discovered equation (and these are highly correlated as shown in Fig.~\ref{fig:corrmatrix}).

\subsubsection{Pressure perturbation}\label{sec:pp}

The second term in Eq.~(\ref{eq:incloudinvariant}) reflects the perturbation pressure field and the associated pressure-gradient force. The importance of drag in thermal momentum budgets has been debated in terms of ``slippery'' \citep{sherwood_slippery_2013} versus ``sticky'' \citep{romps_sticky_2015} thermals, although these studies quantified different terms: momentum dilution by entrainment and the perturbation pressure-gradient force, respectively \citep{morrison2022dynamic}. The two views are therefore not mutually exclusive, and \citet{hernandez-deckers_numerical_2016} also found pressure drag to be the main force opposing buoyancy in simulated cumulus thermals. Our diagnostics address only the pressure-gradient force, and we repeatedly found strong relationships between vertical velocity and pressure perturbation, consistent with a leading-order pressure-gradient force, as in the sticky-thermal view.

The maximum pressure perturbation, $(p'/\bar{p})_{\max}$, is likely a proxy of the dynamic pressure buildup for several reasons. As seen in Figs.~\ref{fig:incloudeqplot} and \ref{fig:corrmatrix}, it scales as $w_{\max}^2$, like we would expect for dynamic pressure. There is also a prominent high pressure perturbation region above the updraft (see Fig.~\ref{fig:composite}c), and the maximum is consistently found there, above the location of $w_{\max}$ (see Fig.~\ref{fig:composite}e).  Our pressure perturbation composite (Fig.~\ref{fig:composite}c) is quite similar to the (absolute) pressure perturbation composite over thermals in Figure 4 of \citet{romps_sticky_2015}.

\begin{figure*}[t]
    \includegraphics{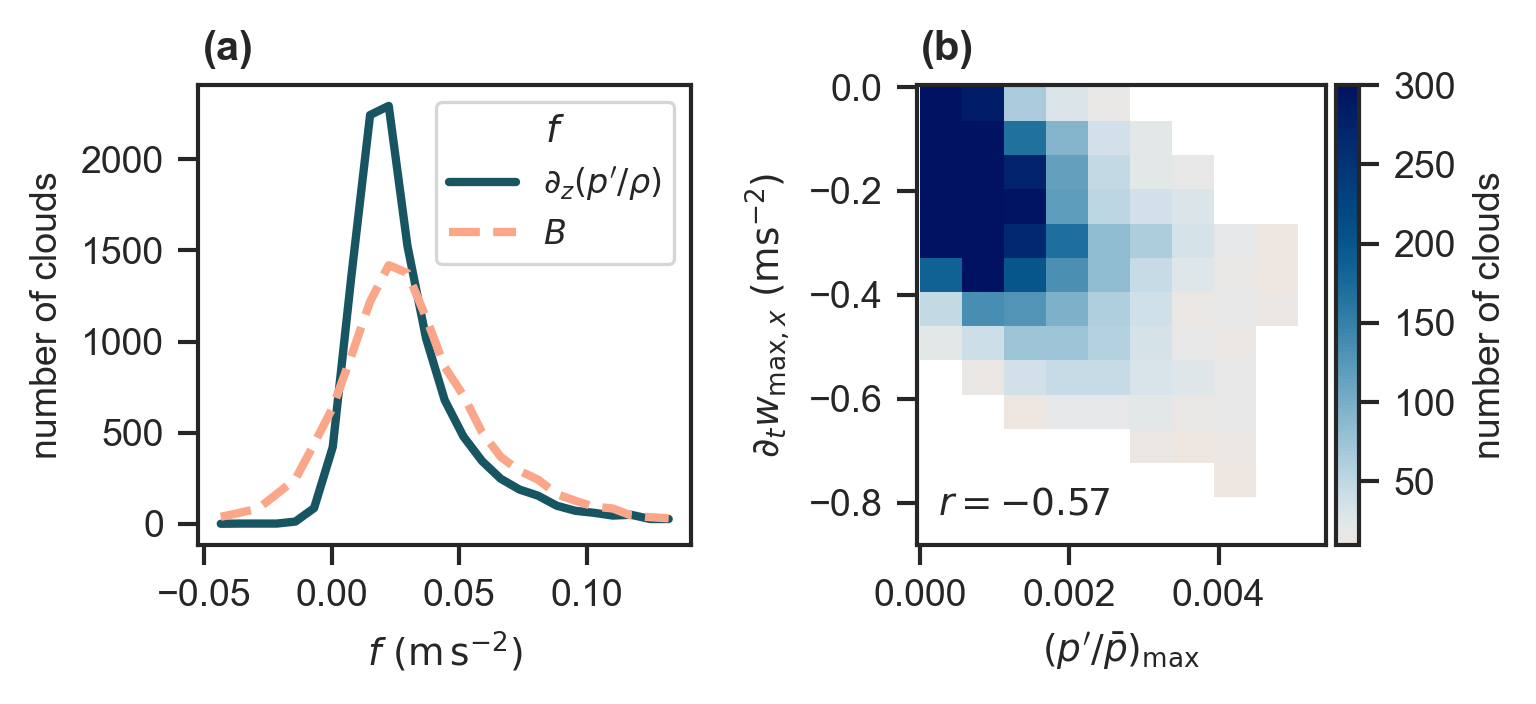}
    \caption{Role of pressure perturbations in updraft dynamics. (a) Histograms of vertical acceleration components diagnosed at the time and location of $w_{\max}$: buoyancy ($B$) and pressure-gradient ($\partial_z(p'/\rho)$) contributions. (b) Relationship between $(p'/\bar{p})_{\max}$ and subsequent deceleration $\partial_t w_{\max ,x}$, where $w_{\max,x}$ is the maximum vertical velocity over the cloud at a given time. $\partial_t w_{\max,x}$ is the slope of a linear fit to $w_{\max,x}$ at the time of peak intensity ($w_{\max,x} = w_{\max}$) and the three following 5-minute outputs. Data from all simulations combined.
    }
    \label{fig:ppd_B_dwdt}
\end{figure*}

The dynamic pressure buildup is driven by high updraft velocity and is associated with a significant pressure-gradient force and deceleration of the updraft. Diagnosing the buoyancy and pressure-gradient forces at the point of $w_{\max}$ reveals they are comparable in magnitude (see Fig.~\ref{fig:ppd_B_dwdt}a), consistent with the sticky-thermal view that the pressure-gradient force and buoyancy are the dominant terms in the vertical momentum budget. A higher $(p'/\overline{p})_{\max}$ at the time of $w_{\max}$ is also correlated with a faster deceleration in the following 15 minutes (see Fig.~\ref{fig:ppd_B_dwdt}b). Here, the deceleration is estimated from the evolution of the cloud-maximum vertical velocity, not at a fixed location, and so reflects the weakening of the cloud's peak updraft rather than of a single air parcel.

These results point to pressure perturbations being a significant term in the vertical momentum balance of deep convective updrafts, rather than a small correction to a buoyancy-dominated balance, consistent with the sticky-thermal view of \citet{romps_sticky_2015} and in contrast to the buoyancy-only assumptions underlying parcel-based theories of updraft intensity.

\subsubsection{Approximate invariant structure and implications for understanding updrafts} \label{sec:incloudeqstructure}

The learned approximate invariant given by Eq.~(\ref{eq:incloudinvariant}) along with the spatial structure of the relevant quantities and their maxima presents a coherent picture of a cloud at the time of peak updraft intensity: condensates and positive temperature anomalies are found in and below the peak ascent velocities and their maxima are proxies for the net effect of moist processes, while dynamic pressure buildup is seen above, with the maximum pressure perturbation indicating the effect of the updraft flow on the pressure field. Although the diagnostics identified here are extrema of cloud-scale quantities, their vertical ordering relative to $w_{\max}$ is systematic and their magnitudes are tightly linked.

Equation~(\ref{eq:incloudinvariant}) has two terms with a physically justifiable scaling ($\sim w_{\max}^2$) separately. However, unlike in the pre-storm equation where there is a clear nonlinear interaction between the two variables, these two have a similar numerical relationship to $w_{\max}$ and along with the third selected feature, the variables are correlated (see Fig.~\ref{fig:corrmatrix}). Thus, the appearance of both $(p'/\bar{p})_{\max}$ and $q_{c,\max}$ together likely reflects a statistical advantage over having only one feature rather than an indication of two independent processes impacting $w_{\max}$.

One possible interpretation of Eq.~(\ref{eq:incloudinvariant}) is that the state of a cloud at the time of $w_{\max}$ is relatively low-dimensional: despite the many processes that shape convective development (entrainment, mixing, microphysics, turbulent variability, \ldots), their integrated effects lead to the peak updraft velocity closely scaling with other in-cloud extrema, with these relationships holding cloud to cloud and across different regimes.

If these scalings prove robust beyond the idealized simulations analyzed here, they may also suggest a pathway for connecting remotely sensed cloud properties to convective intensity. Because quantities related to condensate can be estimated from radar \citep[e.g.,][]{hogan_icewater_2008} or satellite \citep[e.g.,][]{stephens_cloudsat_2008} observations, the strong scaling between condensate extrema and $w_{\max}$ raises the possibility that the observed cloud structure could provide indirect constraints on the peak updraft intensity. At present, however, it remains unclear how well these relationships would generalize to more realistic cloud simulations or to observed convection. An important next step would therefore be to test whether similar scalings emerge in cloud-resolving simulations with more realistic forcing, or observational datasets.

\subsection{Sensitivity to horizontal resolution}\label{sec:resolution_results}

To test whether the discovered relationships hold at higher horizontal resolution than the 1~km grid spacing, we refit Eqs.~(\ref{eq:pysrtwovar}) and~(\ref{eq:incloudpysr}) in the CTRL simulations at 500, 250 and 125~m, both coarsened to 1~km and at native resolution (see Sect.~\ref{sec:resolution_methods}). We compare them with the 1~km CTRL simulation alone ($R^2=0.35$ and $0.86$ for the pre-storm and in-cloud equations, respectively), not with the cross-regime scores of Sects.~\ref{sec:sqrtcape}--\ref{sec:incloudeqstructure}, which include all of the five regimes.

\begin{figure*}[t]
    \includegraphics[width=0.75\textwidth]{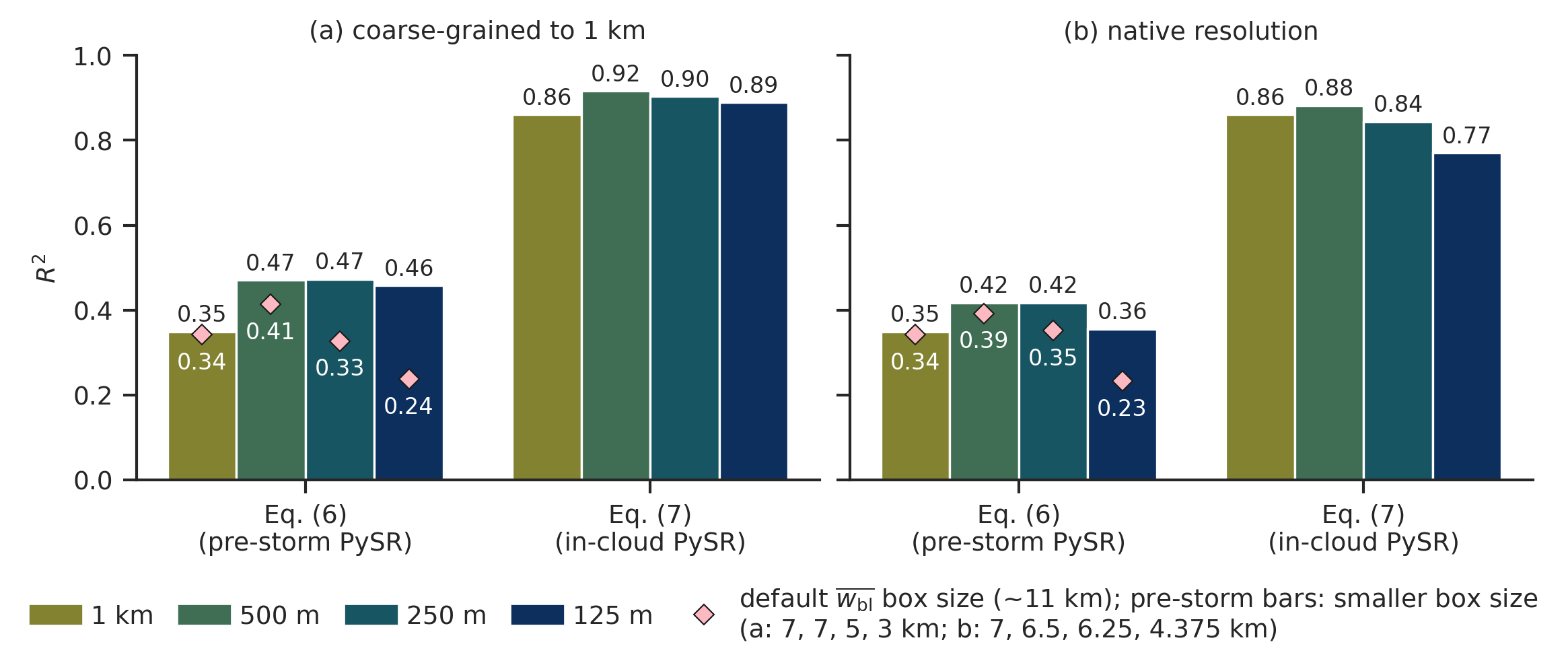}
    \caption{Performance ($R^2$) of the pre-storm (Eq.~(\ref{eq:pysrtwovar})) and in-cloud (Eq.~(\ref{eq:incloudpysr})) equations, refit in the CTRL simulations at 1~km, 500~m, 250~m and 125~m horizontal resolution, with fields (a) coarsened to 1~km and (b) at native resolution. For the pre-storm equation, bars use the $\overline{w_{\mathrm{bl}}}$ box size that maximizes $R^2$ (given in the legend) and diamonds the default box of about 11~km.}
    \label{fig:resolutions}
\end{figure*}

The pre-storm equation (Eq.~(\ref{eq:pysrtwovar})) holds at all resolutions once the size of the box over which $\overline{w_{\mathrm{bl}}}$ is averaged is adapted to the resolution (bars versus diamonds in Fig.~\ref{fig:resolutions}). With the default box of about 11~km, skill decreases with increasing resolution, to $R^2=0.24$ (coarsened) and $0.23$ (native) at 125~m. With the box size chosen to maximize $R^2$ in each simulation, skill matches or exceeds the 1~km value ($R^2=0.46$--$0.47$ coarsened and $0.36$--$0.42$ native). The optimal box shrinks with resolution, from about 7~km at 1~km to 3--4.4~km at 125~m, and does so even when the fields are coarsened to 1~km. The boundary-layer structures associated with $w_{\max}$ therefore become physically smaller at finer resolution, consistent with resolved updrafts not being converged at these grid spacings \citep{jeevanjee2017vertical}. The multiplicative interaction between CAPE and $\overline{w_{\mathrm{bl}}}$ is thus robust to resolution, while the relevant spatial scale of $\overline{w_{\mathrm{bl}}}$ is not.

The in-cloud equation (Eq.~(\ref{eq:incloudpysr})) also remains skillful at all resolutions. With the fields coarsened to 1~km, it explains as much or slightly more of the variance than at 1~km ($R^2=0.89$--$0.92$). At native resolution, its skill is similar at 500 and 250~m and lower at 125~m ($R^2=0.77$), which we attribute to the greater noise of grid-point extrema at finer resolution, because the performance is recovered once the fields are coarsened. Equation~(\ref{eq:incloudpysr}) thus holds at grid spacings where the internal structure of updrafts, including their mixing with the surrounding cloudy air, is better resolved than at 1~km \citep{hernandez-deckers_numerical_2016}, indicating that it is not an artifact of the coarse grid.

\conclusions\label{sec:conclusion}

We identified two compact relationships for maximum updraft velocity ($w_{\max}$) across radiative--convective equilibrium regimes: a predictive pre-storm equation based on CAPE and boundary-layer vertical velocity, and an approximate in-cloud invariant linking $w_{\max}$, maximum pressure perturbation, and maximum cloud condensate at peak intensity. The best pre-storm equation captures 47\% of the variance in $w_{\max}$ ($R^2 = 0.47$, RMSE $= 5.2\ \mathrm{m\,s^{-1}}$), while the best diagnostic in-cloud equation captures 89\% ($R^2 = 0.89$, RMSE $= 2.4\ \mathrm{m\,s^{-1}}$). Both equations remain skillful in simulations at horizontal grid spacings down to 125~m, with the relevant averaging scale of $\overline{w_{\mathrm{bl}}}$ decreasing with resolution. The results show that combining feature selection validated across regimes with equation discovery can recover physically meaningful, generalizable relationships, illustrating the potential of equation discovery approaches for advancing process understanding beyond prediction alone.

For pre-storm prediction, CAPE is the most informative single predictor across regimes, consistent with its widespread use in convective forecasting and its role as a theoretical upper bound on updraft kinetic energy. The symbolic regression algorithm independently rediscovered the classical $w_{\max} \sim \sqrt{\mathrm{CAPE}}$ scaling (Eq.~(\ref{eq:capescaling})), providing methodological validation. However, CAPE alone explains only $\approx$25\% of the cross-regime variance and has no predictive value within a single simulation ($R^2 \approx 0$). This failure is consistent with the recognized limitation that CAPE characterizes the thermodynamic potential of the environment but does not capture the variability in in-cloud processes that determines how much of that potential is realized \citep{parodi_theory_2009, singh_increases_2015}. The mean boundary layer vertical velocity $\overline{w_{\mathrm{bl}}}$, measured 30 minutes before $w_{\max}$, is essential for capturing some cloud-to-cloud variability and interacts nonlinearly with CAPE, suppressing $w_{\max}$ even in high-CAPE environments. This nonlinear interaction is not captured by simply adding $\overline{w_{\mathrm{bl}}}$ as an initial kinetic energy term: the baseline equation (Eq.~(\ref{eq:capetrigger})) yields $R^2 = 0.32$, barely improving on CAPE alone ($R^2 = 0.25$) despite adding a second variable. Boundary layer dynamics therefore play a larger role than simply setting the initial kinetic energy of an updraft, as demonstrated by the nonlinear interaction in the learned equation (Eq.~(\ref{eq:pysrtwovar})) resulting in $R^2 = 0.47$. 

For in-cloud diagnostics, the maximum pressure perturbation $(p'/\bar{p})_{\max}$ and maximum cloud condensate $q_{c,\max}$ together capture 89\% of the variance in $w_{\max}$ through Eq.~(\ref{eq:incloudpysr}). The selection of $(p'/\bar{p})_{\max}$ and the comparable magnitudes of the pressure-gradient force and buoyancy at peak intensity (Fig.~\ref{fig:ppd_B_dwdt}) are consistent with the sticky-thermal view \citep{romps_sticky_2015}, in which the pressure-gradient force is a significant term in the vertical momentum budget alongside buoyancy. However, we diagnose only this pressure-gradient force, not the dilution of updraft momentum by entrainment emphasized by \citet{sherwood_slippery_2013}, and at 1~km horizontal grid spacing the internal circulation of individual thermals is not well resolved, so our results do not directly test the slippery-thermal mechanism. The role of $q_{c,\max}$ as a proxy for latent heating is consistent with the expected tight coupling between condensation rate and vertical velocity \citep{grant_linear_2022}, although the scaling here is over the whole-cloud extrema rather than a local instantaneous relationship and involves the cloud condensate rather than its time derivative. The co-selection of both $q_{c,\max}$ and $(p'/\bar{p})_{\max}$ likely reflects a statistical advantage over either predictor alone rather than two independent physical processes. It is also worth noting that the in-cloud invariant is strictly diagnostic rather than predictive: its equation describes the cloud state at peak intensity, relating several in-cloud extrema, rather than describing a particular causal sequence.

A key open question is the physical interpretation of $\overline{w_{\mathrm{bl}}}$, which may reflect cold-pool convergence and lifting \citep{fuglestvedt_coldpools_2020, tompkins_coldpools_2001, torri_mechanisms_2015}, the net upward boundary layer mass flux, the geometry of the developing updraft, or warm or moist anomalies in its inflow (Sect.~\ref{sec:whatiswbl}). Its lag sensitivity (the cross-validation $R^2$ of the CAPE and $\overline{w_{\mathrm{bl}}}$ model drops from 0.45 to 0.34 when $\overline{w_{\mathrm{bl}}}$ is measured 45 rather than 30 minutes before $w_{\max}$) and the fact that conditioning on updraft or cloud grid points worsens performance both point to a more complex role than a simple triggering velocity. Some of the remaining unexplained pre-storm variance is recovered by diagnosing the height at which $w_{\max}$ occurs: using CAPE integrated to the correct $z_{\max}$ nearly doubles the explained variance from $R^2 = 0.25$ for CAPE integrated to the equilibrium level to $R^2 = 0.49$ (Sect.~\ref{sec:zmax}), pointing to the importance of processes that determine where vertical acceleration stops within a cloud. The remaining unexplained variance likely reflects a combination of inherent cloud-to-cloud stochasticity and processes not captured by any of our predictors.

These results carry several important limitations. All simulations are idealized RCE over a fixed sea surface temperature without large-scale wind shear, moisture variability, or organized convective forcing. The influence of shear on updraft dynamics and associated pressure perturbations \citep{hernandez-deckers_numerical_2016, leger_simple_2019} is absent, and how the CAPE--$\overline{w_{\mathrm{bl}}}$ interaction generalizes to sheared environments or over land remains untested. Although the features were selected and validated on simulations withheld from training, they are constrained by five simulations, and their validity outside the sampled range of SST and radiative cooling remains to be tested. The resolution tests (Sect.~\ref{sec:resolution_results}) cover only the CTRL regime and evaluate the equations found at 1~km without repeating the feature selection. Moreover, the averaging scale of $\overline{w_{\mathrm{bl}}}$ that maximizes skill depends on resolution, so applying the pre-storm equation in other models requires choosing this scale accordingly. Another limitation is that domain-mean precipitation in RCE is constrained by radiative cooling, while similarly sized tropical regions are usually far from RCE and so their precipitation can vary much more due to atmospheric circulations \citep{jakob_radiative_2019,muller2011energetic}. Our simulations span a range of thermodynamic regimes, but they do not capture this circulation-driven variability. At the same time, this simplified setting is what allowed us to sample a wide range of climates and isolate the processes regulating $w_{\max}$.

Understanding what governs convective updraft velocity has direct relevance for how deep convection behaves across different climate states. Updraft velocities control ice lofting and lightning flash rates \citep{romps_evaluating_2019}, and changes in updraft intensity under warming modulate precipitation extremes through dynamical effects not fully captured by thermodynamic scaling alone \citep{muller_extremes_2020, ogorman_schneider_2009}. The finding that boundary layer dynamics regulate $w_{\max}$ beyond what CAPE predicts, and that $w_{\max}$ is tightly linked to in-cloud processes, suggests that these dynamical contributions may respond to warming differently from the thermodynamic environment, with implications for how convective intensity changes across climate states.

Equations derived here, which generalize across our RCE regimes, offer a physically interpretable basis for understanding what sets convective intensity across regimes and at the individual cloud level, and for forming testable hypotheses about which processes drive that variability. They are intended for process understanding rather than direct use in cumulus parameterizations, as they rely on cloud-scale diagnostics, such as per-cloud extrema, that coarse-resolution models do not resolve. They may nonetheless inform parameterization design indirectly, for example through the finding that boundary-layer vertical velocity modulates $w_{\max}$ nonlinearly rather than as an additive kinetic energy term. Because some of the identified scalings depend on condensate-related quantities that can be retrieved from radar or satellite observations \citep{hogan_icewater_2008,stephens_cloudsat_2008}, these relationships may also provide a pathway for connecting observed cloud structure to convective intensity.

More broadly, the idealized RCE framework should be viewed as a foundation rather than an endpoint: having identified a compact set of variables that captures $w_{\max}$ across a wide range of climates, the methodology can now be applied to global storm-resolving models (GSRMs), where large-scale forcing, wind shear, and convective organization play a larger role. In GSRMs, updrafts are explicit but not converged at grid spacings comparable to or coarser than ours \citep{hohenegger2020climate, jeevanjee2017vertical}, and $w_{\max}$ is used directly in some schemes, e.g., for lightning \citep{barthe2010estimation}. While the fitted coefficients are specific to our set-up, the low-dimensional spaces identified here (the CAPE--$\overline{w_{\mathrm{bl}}}$ plane and the approximate invariant at peak intensity) can serve as process-oriented diagnostics for comparing GSRMs across models, resolutions and climate regimes. Such simulations sample a narrower range of climates but a much broader range of forcing regimes, making them complementary to the present study and an important next step for testing the generality of these results and identifying which additional processes regulate convective intensity in more realistic settings. This is timely, as storm-resolving simulations of warmer climates are increasingly being performed \citep[e.g.,][]{cheng_2022_impact, bolot_kilometer-scale_2023, segura_2025_nextgems}. For those climates, no observational ground truth exists, and such physically grounded diagnostics offer a way to evaluate and interpret simulated changes in convective intensity.

\codedataavailability{The code used in this study, the processed cloud-level dataset and SAM simulation configuration files underlying all analysis and figures in this study are archived at Zenodo: \url{https://doi.org/10.5281/zenodo.21474646} \citep{pechacova_2026_zenodo}. The System for Atmospheric Modeling (SAM) source code is available from \url{http://rossby.msrc.sunysb.edu/SAM.html}. The raw SAM simulation output is available from the corresponding author upon reasonable request, subject to storage constraints.}

\appendix

\section{Feature lists and Pareto fronts}

\begin{figure*}[t]
\includegraphics[width=12cm]{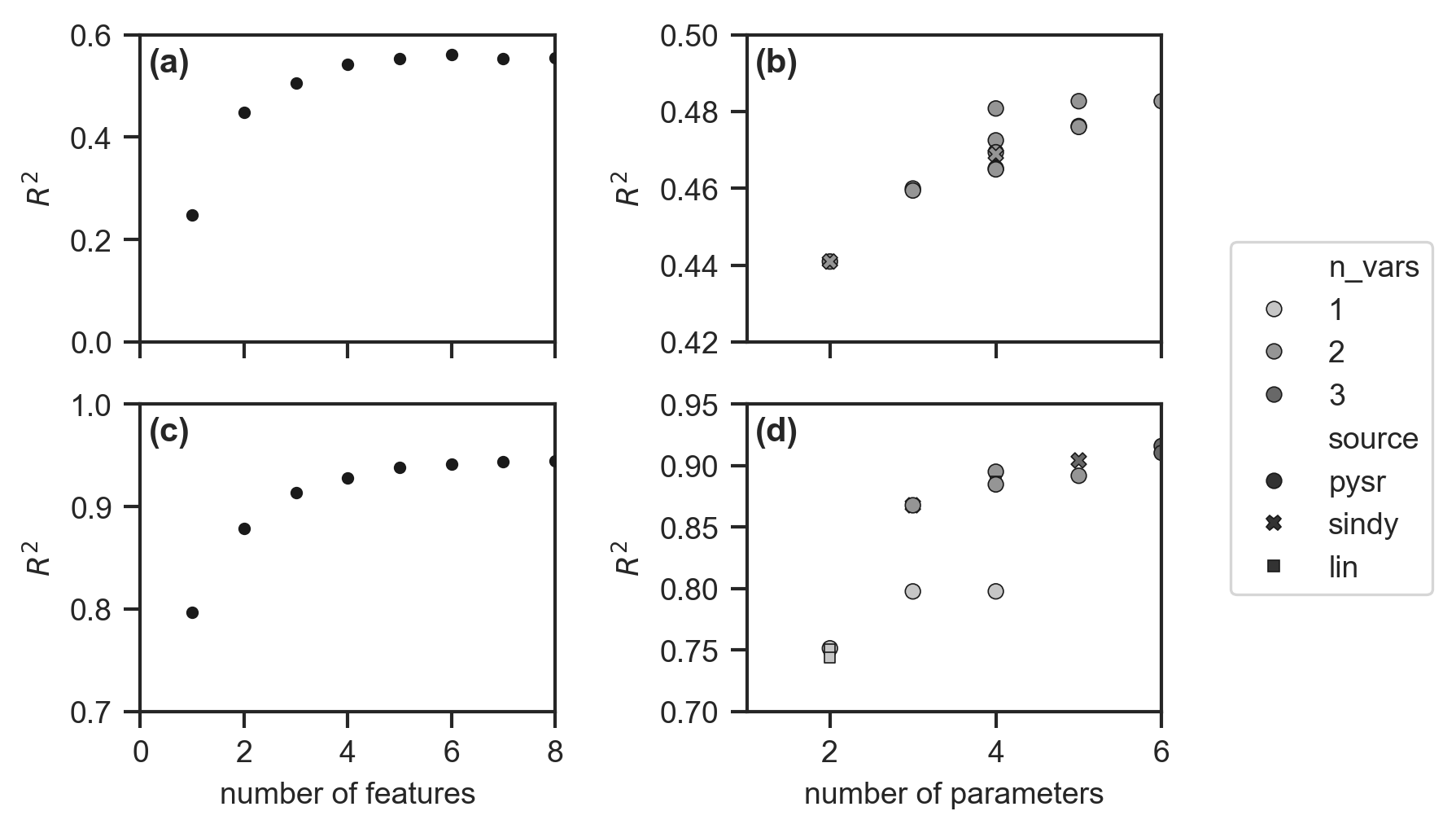}
\caption{
Pareto fronts from sequential feature selection. (a,c) Predictive performance ($R^2$) of neural-network models using (a) pre-storm predictors and (c) in-cloud diagnostics as a function of the number of selected features. Each point corresponds to a model trained during feature selection. (b,d) Performance of discovered equations as a function of the number of parameters.
}
\label{fig:prestorm_and_incloud_paretos}
\end{figure*}

See Tables~\ref{table:prestorm} and \ref{table:incloud} for lists of pre-storm and in-cloud features, respectively, supplementing the descriptions in Sects.~\ref{sec:prestorm_feature_defs} and \ref{sec:incloud_feature_defs}. Figure~\ref{fig:prestorm_and_incloud_paretos} shows the Pareto fronts with feature selection and equation learning results.

\appendixtables
\begin{table*}[t]
\caption{List of pre-storm  features (calculated over all points in the 11$\times$11~km square around the location of $w_{\max}$, lagged by 0.5 to 2 hours relative to the time of $w_{\max}$).}
\begin{tabular}{p{4cm}cp{5cm}p{3.75cm}}
\tophline
Vertical location & Symbol & Description & Computation method(s) \\
\middlehline
Vertical integrals & $\mathrm{CAPE}_{\hat{\epsilon} X}$ and $\mathrm{CIN_{\hat{\epsilon} X}}$ & Convective available potential energy and convective inhibition, computed from the vertically integrated buoyancy of an entraining plume following \citet{singh_influence_2013}, with an entrainment profile given by $\epsilon (z) = \hat{\epsilon}/z$ with constant $\hat{\epsilon} = 0,~0.5,~1.0$ & See Sect.~\ref{sec:prestorm_feature_defs} for details. \\
& $\mathrm{CRH}$ & Column relative humidity & surface to tropopause ($200~\mathrm{K}$) \\
\middlehline
\multirow{8}{*}{Boundary layer (0--1~km)} & $C_\mathrm{bl}$ & Horizontal convergence  & \\
\cline{3-4}
 & $\mathrm{RH}_\mathrm{bl}$ & Relative humidity & \\
& $w_\mathrm{bl}$ & Vertical velocity & \\
& $T'_\mathrm{bl}$ & Temperature perturbation relative to the domain-mean profile &  \\
& $V_\mathrm{bl}$ & Horizontal wind magnitude & \\
\cline{1-3}
Free troposphere (1~km to tropopause (200~K level)) & $\mathrm{RH}_\mathrm{ft}$ & Relative humidity & mean, standard deviation, minimum, maximum (all volume-weighted to account for non-uniform vertical spacing)\\
& $T'_\mathrm{ft}$ & Temperature perturbation relative to the domain-mean profile & \\
& $V_{260\mathrm{K}}$ & Horizontal wind magnitude at the height where the domain-mean temperature is $260 \mathrm{~K}$ & \\
& $V_{230\mathrm{K}}$ & Horizontal wind magnitude at the height where the domain-mean temperature is $230 \mathrm{~K}$ & \\

\bottomhline
\label{table:prestorm}

\end{tabular}
\end{table*}

\begin{table*}[t]
\caption{List of in-cloud  features. See Sect.~\ref{sec:incloud_feature_defs} for more detail.}
\begin{tabular}{p{4cm}cp{5cm}p{3.75cm}}
\tophline
Location/mask & Symbol & Description &  Spatial statistics \\

\middlehline

Cloud mask (defined by tracking algorithm) & $v_T$ & Terminal velocity of hydrometeors &  \\
& $q_c$ & Cloud condensate mixing ratio &  \\
& $T'$ & Temperature perturbation relative to the domain-mean profile & \\
& $B$ & Buoyancy (calculated using $B \approx g(T'/\overline{T}+0.608~q_v' - q_c-p'/\overline{p})$ following~\citet{khairoutdinov_cloud_2003}) & \\
& $\frac{p'}{\overline{p}}$ & Fractional pressure perturbation relative to and normalized by the domain-mean profile ($\overline{p}$) & \\

Environment mask (defined as an 11$\times$11~km square around the location of $w_{\max}$, excluding any cloud regions), from the level of free convection up to the tropopause (200~K level) & $\mathrm{RH}_{\mathrm{env}}$ & Relative humidity & mean, standard deviation, percentiles: 0, 20, 50, 80, 100 (all volume-weighted to account for non-uniform vertical spacing)\\
 & $w_{\mathrm{env}}$ & Vertical velocity & \\
& $T'_{\mathrm{env}}$ & Temperature perturbation relative to the domain-mean profile & \\
& $q_{v,\mathrm{env}}'$ & Water vapor mixing ratio perturbation relative to the domain-mean profile & \\
& $V_{260\mathrm{~K, env}}$ & Horizontal wind magnitude at the height where the domain-mean temperature is $260 \mathrm{~K}$ & \\
& $V_{230\mathrm{~K, env}}$ & Horizontal wind magnitude at the height where the domain-mean temperature is $230 \mathrm{~K}$ & \\

\bottomhline
\label{table:incloud}

\end{tabular}
\end{table*}

\begin{table*}[t]
\caption{Sensitivity of CAPE-based predictions of $w_{\max}$ to the entrainment parameter $\hat\epsilon$. The CAPE ratio is the range over the five simulations of simulation-mean CAPE with a given $\hat\epsilon$ relative to $\hat\epsilon = 0$, and $r$ is the Pearson correlation with CAPE for $\hat\epsilon = 0$ across all clouds. Neural network (NN) $R^2$ values are from the LOGO cross-validation with CAPE as the only feature. For Eq.~(\ref{eq:capescaling}), $a_0$ and $w_0$ are fitted to all clouds, $R^2$ is computed over all clouds, and $R^2_{\mathrm{means}}$ compares the simulation-mean predictions with the five simulation-mean values of $w_{\max}$.}
\label{table:epsilon}
\begin{tabular}{lcccccccccc}
\tophline
$\hat\epsilon$ & CAPE ratio & $r$ & NN $R^2_{\mathrm{across}}$ & NN $R^2_{\mathrm{within}}$ & $a_0$ & $w_0$ (m~s$^{-1}$) & Eq.~(\ref{eq:capescaling}) $R^2$ & Eq.~(\ref{eq:capescaling}) $R^2_{\mathrm{means}}$ \\
\middlehline
0   & 1         & 1     & 0.21 & $-0.10$ & 0.17 & 10.1 & 0.22 & 0.94 \\
0.5 & 0.58--0.60 & 0.996 & 0.23 & $-0.05$ & 0.23 & 9.9 & 0.24 & 0.94 \\
1.0 & 0.23--0.27 & 0.952  & 0.25 & $-0.02$ & 0.33 & 10.4 & 0.25 & 0.94 \\
\bottomhline
\end{tabular}
\end{table*}

\section{Diagnosing the sources of CAPE variance}\label{appendix:capevariance}

The diagnostic relationship for $w_{\max}$ with CAPE with the correct integration limit, discussed in Sect.~\ref{sec:zmax} can be written as

\begin{equation}\label{eq:capeintegral}
    w_{\max}^2 \approx 2 \times \mathrm{CAPE}_{z_{\max}} = 2\times \int_{z_{\mathrm{LFC}}}^{z_{{\max}}} 
g \, \frac{T_{\mathrm{parcel}} - T_{\mathrm{env}}}{T_{\mathrm{env}}} \, dz
\end{equation}

where $T_{\mathrm{parcel}}$ and $T_{\mathrm{env}}$ are the temperatures of an idealized parcel trajectory, and the environment, respectively. $z_{\mathrm{LFC}}$ and $z_{{\max}}$ are the heights of the level of free convection and of the location where $w_{{\max}}$ occurs. This is a function of many variables and variations in the integration limits (mainly $z_{{\max}}$) as well as the integrand (initial parcel properties, environmental temperature, possibly humidity if we account for entrainment, etc.) will contribute to the variations in $w_{\max}$. 

Approximating the integrand in Eq.~(\ref{eq:capeintegral}) as a linear profile, we can write CAPE integrated up to $z_{\max}$ as 

\begin{equation}\label{eq:capezmaxproduct}
    \mathrm{CAPE}_{z_{\max}} \approx \mathrm{CAPE~} \times \frac{z_{\max}^2 - z_{\mathrm{LFC}}^2}{z_{\mathrm{p}}^2 - z_{\mathrm{LFC}}^2}
\end{equation}

where $z_{\mathrm{p}}$ is a (simulation-specific) constant. This linear profile approximation works well for $\hat\epsilon=0, 0.5$ (at $\hat\epsilon=1$, the profiles vary in shape too substantially) and is relatively insensitive to the choice of $z_\mathrm{p}$. Having a simple product of two terms, we can decompose the contributions of the variations of CAPE and $z_{\max}$ to the variations of $\mathrm{CAPE}_{z_{\max}}$:

\begin{equation}\label{eq:capevariance}
    \mathrm{Var}(\log\mathrm{CAPE}_{z_{\max}}) \approx \mathrm{Var}(\log\mathrm{CAPE}) +  \mathrm{Var}\left(\log{\frac{z_{\max}^2 - z_{\mathrm{LFC}}^2}{z_{\mathrm{p}}^2 - z_{\mathrm{LFC}}^2}}\right)
\end{equation}

\begin{figure}[t]
    \centering
    \includegraphics{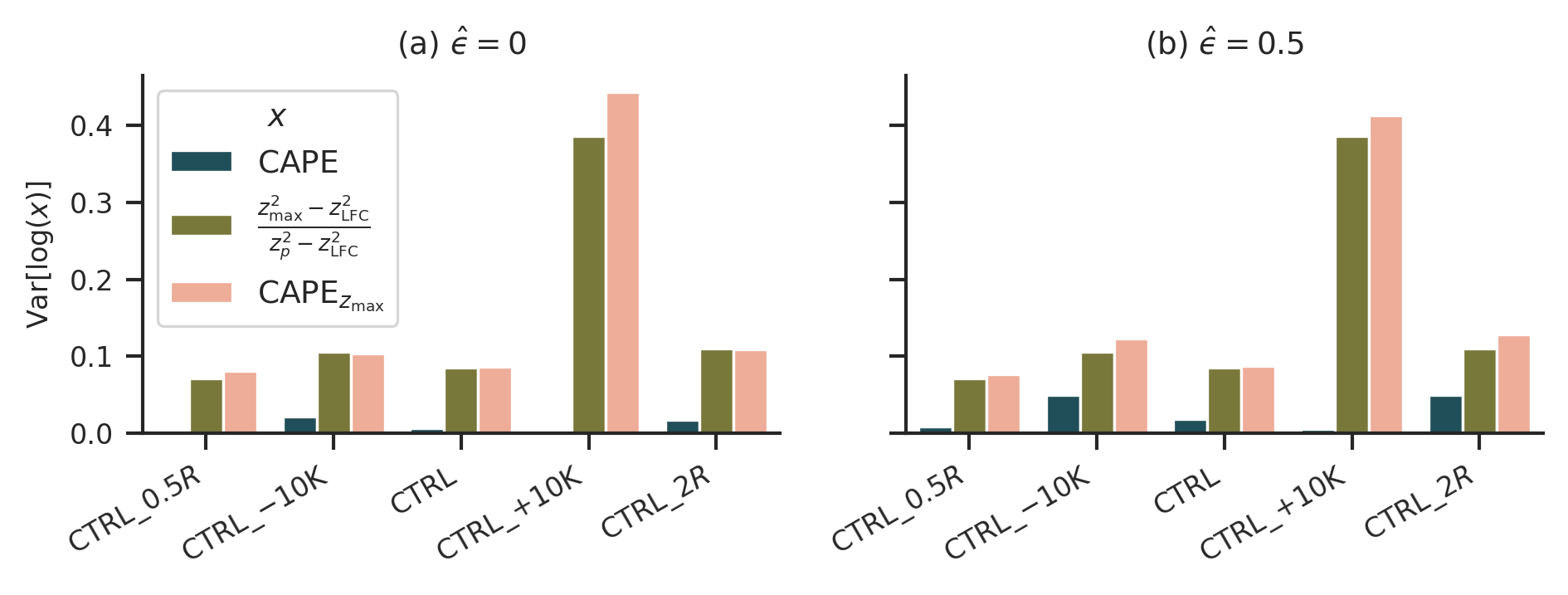}
    \caption{Decomposition of variance in $\mathrm{CAPE}_{z_{\max}}$. Contributions from variability in $\mathrm{CAPE}$ and in the effective integration limit (related to $z_{\max}$) diagnosed using the approximation in Eq.~(\ref{eq:capevariance}) are shown for (a) $\hat\epsilon=0$ and (b) $\hat\epsilon=0.5$.}
    \label{fig:capevariances}
\end{figure}

Figure~\ref{fig:capevariances} shows these contributions in our dataset for $\hat\epsilon=0$ and $0.5$ and the second term dominates over the first term in all of them, which is largely due to a relatively homogeneous environment in our set-up. For $\hat\epsilon=1$, the linear profile assumption allowing us to write Eq.~(\ref{eq:capezmaxproduct}) is no longer valid so the figure is not shown.

\section{Equation coefficient robustness}\label{appendix:eqrobustness}

Results of the LOGO coefficient refitting described in Sect.~\ref{sec:eq_evaluation} are shown in Tables~\ref{tab:logo_prestorm} and~\ref{tab:logo_incloud}. Refitting the coefficients on four simulations and predicting the held-out fifth changes the RMSE only modestly for both equations, indicating that neither equation is overfit to a particular regime. For the pre-storm equation (Eq.~(\ref{eq:pysrtwovar})), the held-out RMSE increases by under 5\% for every simulation except the coldest. The in-cloud equation (Eq.~(\ref{eq:incloudpysr})) behaves similarly, with increases up to 10\% outside the warmest case. The largest degradations are less than 25\% for both cases and even for those, the absolute RMSE stays small.

We calculated the coefficient uncertainties quoted in the main text with a bootstrap, drawing 1000 resamples of clouds stratified by simulation with replacement and refitting the coefficients on each resample. We then report the standard deviation of each coefficient, and these values reflect sampling noise given the five simulations, complementing the across-simulation sensitivity captured by the LOGO refitting.

\begin{table}[h]
\centering
\caption{LOGO cross-validation results for the pre-storm equation (Eq.~(\ref{eq:pysrtwovar})). $\sigma_{w_{\max}}$ denotes the standard deviation of $w_{\max}$ in the held-out simulation, RMSE$_{\mathrm{full}}$ uses coefficients fitted on all five simulations, and RMSE$_{\mathrm{LOGO}}$ uses coefficients fitted on the remaining four.}
\label{tab:logo_prestorm}
\begin{tabular}{lccc}
\hline
Simulation & $\sigma_{w_{\max}}$ & RMSE$_{\mathrm{full}}$ & RMSE$_{\mathrm{LOGO}}$ \\
           & (m\,s$^{-1}$)       & (m\,s$^{-1}$)          & (m\,s$^{-1}$)          \\
\hline
CTRL            & 5.2 & 4.6 & 4.6 \\
CTRL\_2$R$      & 4.6 & 4.1 & 4.2 \\
CTRL\_0.5$R$    & 5.9 & 5.1 & 5.2 \\
CTRL\_--10K     & 3.1 & 3.2 & 3.8 \\
CTRL\_+10K      & 10.2 & 7.8 & 8.1 \\
\hline
Mean            & 5.8 & 5.0 & 5.2 \\
\hline
\end{tabular}
\end{table}

\begin{table}[h]
\centering
\caption{As Table~\ref{tab:logo_prestorm}, but for the in-cloud equation (Eq.~(\ref{eq:incloudpysr})).}
\label{tab:logo_incloud}
\begin{tabular}{lccc}
\hline
Simulation & $\sigma_{w_{\max}}$ & RMSE$_{\mathrm{full}}$ & RMSE$_{\mathrm{LOGO}}$ \\
           & (m\,s$^{-1}$)       & (m\,s$^{-1}$)          & (m\,s$^{-1}$)          \\
\hline
CTRL            & 5.2 & 2.2 & 2.2 \\
CTRL\_2$R$      & 4.7 & 2.1 & 2.1 \\
CTRL\_0.5$R$    & 5.9 & 2.2 & 2.2 \\
CTRL\_--10K     & 3.1 & 1.5 & 1.6 \\
CTRL\_+10K      & 10.2 & 3.4 & 4.2 \\
\hline
Mean            & 5.8 & 2.3 & 2.5 \\
\hline
\end{tabular}
\end{table}

\noappendix

\authorcontribution{
Conceptualization: BP, AC, CM. 
Methodology: BP, TB, CM, AC. 
Software: BP, LA, AC. 
Formal analysis: BP, TB, CM, LA, AC. 
Investigation: BP. 
Data curation: BP. 
Writing -- original draft preparation: BP. 
Writing -- review and editing: AC, LA, TB, CM. 
Visualization: BP. 
Supervision: CM, TB. 
Validation: BP. 
}

\competinginterests{The authors declare that they have no conflict of interest.} 

\begin{acknowledgements}
    We thank Andrea Polesello for a helpful discussion of the CAPE variance analysis in Appendix~\ref{appendix:capevariance}, Francesco Locatello for insightful discussions that helped shape this work, Steven Sherwood for useful discussions and feedback, and Jiawei Bao for helpful discussions and providing the code for the entraining plume buoyancy calculation. We are also grateful to Martin Singh and an anonymous reviewer for their constructive comments, which substantially improved the manuscript. This project has received funding from the European Union’s Horizon 2020 research and innovation programme under the Marie Skłodowska-Curie grant agreement No 101034413 awarded to AC. TB acknowledges support from the Swiss State Secretariat for Education, Research, and Innovation (SERI) under the Horizon Europe AI4PEX Project (Grant Agreement ID 101137682; SERI No. 23.00546). The lightbulb icon in Fig.~\ref{fig:methods} was generated using ChatGPT and subsequently modified by hand.
\end{acknowledgements}

\bibliographystyle{copernicus}
\bibliography{references.bib}

\begin{thebibliography}{66}
\providecommand{\natexlab}[1]{#1}
\providecommand{\url}[1]{\texttt{#1}}
\providecommand{\urlprefix}{}
\expandafter\ifx\csname urlstyle\endcsname\relax
  \providecommand{\doi}[1]{https://doi.org/\discretionary{}{}{}#1}\else
  \providecommand{\doi}{https://doi.org/\discretionary{}{}{}\begingroup \urlstyle{rm}\Url}\fi

\bibitem[{Abramian et~al.(2023)Abramian, Muller, and Risi}]{abramian_extreme_2023}
Abramian, S., Muller, C., and Risi, C.: Extreme Precipitation in Tropical Squall Lines, Journal of Advances in Modeling Earth Systems, 15, e2022MS003\,477, \doi{10.1029/2022MS003477}, 2023.

\bibitem[{Agasthya and Muller(2025)}]{agasthya_moist_2025}
Agasthya, L. and Muller, C.: Moist convection and radiative cooling: Dynamical response and scaling, Quarterly Journal of the Royal Meteorological Society, 152, e70\,044, \doi{10.1002/qj.70044}, 2025.

\bibitem[{Agasthya et~al.(2025)Agasthya, Muller, and Cheve}]{agasthya_scaling_2024}
Agasthya, L., Muller, C., and Cheve, M.: Moist convective scaling: Insights from an idealised model, Quarterly Journal of the Royal Meteorological Society, 151, e4902, \doi{10.1002/qj.4902}, 2025.

\bibitem[{Barthe et~al.(2010)Barthe, Deierling, and Barth}]{barthe2010estimation}
Barthe, C., Deierling, W., and Barth, M.~C.: Estimation of total lightning from various storm parameters: A cloud-resolving model study, Journal of Geophysical Research: Atmospheres, 115, 2010.

\bibitem[{Beucler et~al.(2024)Beucler, Gentine, Yuval, Gupta, Peng, Lin, Yu, Rasp, Ahmed, O’Gorman, Neelin, Lutsko, and Pritchard}]{beucler2024climateinvariant}
Beucler, T., Gentine, P., Yuval, J., Gupta, A., Peng, L., Lin, J., Yu, S., Rasp, S., Ahmed, F., O’Gorman, P.~A., Neelin, J.~D., Lutsko, N.~J., and Pritchard, M.: Climate-invariant machine learning, Science Advances, 10, eadj7250, \doi{10.1126/sciadv.adj7250}, 2024.

\bibitem[{Beucler et~al.(2025)Beucler, Grundner, Shamekh, Ukkonen, Chantry, and Lagerquist}]{beucler_distilling_2024}
Beucler, T., Grundner, A., Shamekh, S., Ukkonen, P., Chantry, M., and Lagerquist, R.: Distilling Machine Learning’s Added Value: Pareto Fronts in Atmospheric Applications, Artificial Intelligence for the Earth Systems, 4, e240\,078, \doi{10.1175/AIES-D-24-0078.1}, 2025.

\bibitem[{Bolot et~al.(2023)Bolot, Harris, Cheng, Merlis, Blossey, Bretherton, Clark, Kaltenbaugh, Zhou, and Fueglistaler}]{bolot_kilometer-scale_2023}
Bolot, M., Harris, L.~M., Cheng, K.-Y., Merlis, T.~M., Blossey, P.~N., Bretherton, C.~S., Clark, S.~K., Kaltenbaugh, A., Zhou, L., and Fueglistaler, S.: Kilometer-scale global warming simulations and active sensors reveal changes in tropical deep convection, 6, 1--8, \doi{10.1038/s41612-023-00525-w}, publisher: Nature Publishing Group, 2023.

\bibitem[{Bony et~al.(2016)Bony, Stevens, Coppin, Becker, Reed, Voigt, and Medeiros}]{bony_anvil_2016}
Bony, S., Stevens, B., Coppin, D., Becker, T., Reed, K.~A., Voigt, A., and Medeiros, B.: Thermodynamic control of anvil cloud amount, Proceedings of the National Academy of Sciences, 113, 8927--8932, \doi{10.1073/pnas.1601472113}, 2016.

\bibitem[{Bony et~al.(2025)Bony, Poujol, McKim, Rochetin, Lothon, Windmiller, Maury, Dufaux, Jaffeux, Chazette, and Delano\"e}]{bonyRoleThermals2025}
Bony, S., Poujol, B., McKim, B., Rochetin, N., Lothon, M., Windmiller, J., Maury, N., Dufaux, C., Jaffeux, L., Chazette, P., and Delano\"e, J.: Evidence for the role of thermal and cloud merging in mesoscale convective organization, Atmospheric Chemistry and Physics, 25, 17\,331--17\,362, \doi{10.5194/acp-25-17331-2025}, 2025.

\bibitem[{Brunton et~al.(2016)Brunton, Proctor, and Kutz}]{brunton_discovering_2016}
Brunton, S.~L., Proctor, J.~L., and Kutz, J.~N.: Discovering governing equations from data by sparse identification of nonlinear dynamical systems, Proceedings of the National Academy of Sciences, 113, 3932--3937, \doi{10.1073/pnas.1517384113}, 2016.

\bibitem[{Casallas et~al.(2024)Casallas, Cabrera, Guevara-Luna, Tompkins, Gonzalez, Aranda, Belalcázar, Mogollon-Sotelo, Celis, Lopez-Barrera, Pena-Rincon, and Ferro}]{casallas_airpollution_2024}
Casallas, A., Cabrera, A., Guevara-Luna, M., Tompkins, A., Gonzalez, Y., Aranda, J., Belalcázar, L., Mogollon-Sotelo, C., Celis, N., Lopez-Barrera, E., Pena-Rincon, C., and Ferro, C.: Air Pollution Analysis in Northwestern South America: A New Lagrangian Framework, Science of the Total Environment, 906, 167\,350, \doi{10.1016/j.scitotenv.2023.167350}, 2024.

\bibitem[{Casallas et~al.(2025)Casallas, Tompkins, Muller, and Thompson}]{casallas_sensitivity_2025}
Casallas, A., Tompkins, A.~M., Muller, C., and Thompson, G.: Sensitivity of Self-Aggregation and the Key Role of the Free Convection Distance, Journal of Advances in Modeling Earth Systems, 17, e2024MS004\,791, \doi{10.1029/2024MS004791}, 2025.

\bibitem[{Chen and Guestrin(2016)}]{chen_xgboost_2016}
Chen, T. and Guestrin, C.: {XGBoost}: A Scalable Tree Boosting System, in: Proceedings of the 22nd ACM SIGKDD International Conference on Knowledge Discovery and Data Mining, KDD '16, pp. 785--794, ACM, New York, NY, USA, ISBN 978-1-4503-4232-2, \doi{10.1145/2939672.2939785}, 2016.

\bibitem[{Cheng et~al.(2022)Cheng, Harris, Bretherton, Merlis, Bolot, Zhou, Kaltenbaugh, Clark, and Fueglistaler}]{cheng_2022_impact}
Cheng, K.-Y., Harris, L., Bretherton, C., Merlis, T.~M., Bolot, M., Zhou, L., Kaltenbaugh, A., Clark, S., and Fueglistaler, S.: Impact of warmer sea surface temperature on the global pattern of intense convection: Insights from a global storm resolving model, Geophysical Research Letters, 49, e2022GL099\,796, 2022.

\bibitem[{Cranmer(2023)}]{cranmer_pysr_2023}
Cranmer, M.: Interpretable Machine Learning for Science with PySR and SymbolicRegression.jl, \urlprefix\url{https://arxiv.org/abs/2305.01582}, 2023.

\bibitem[{Emanuel(1994)}]{Emanuel1994}
Emanuel, K.: {Atmospheric Convection}, {Oxford University Press}, New York, USA, 1994.

\bibitem[{Forster et~al.(2021)Forster, Storelvmo, Armour, Collins, Dufresne, Frame, and Zhang}]{ipcc_energybudget_2021}
Forster, P., Storelvmo, T., Armour, K., Collins, W., Dufresne, J.-L., Frame, D., and Zhang, H.: The Earth's Energy Budget, Climate Feedbacks, and Climate Sensitivity, in: IPCC Sixth Assessment Report, pp. 923--1054, Cambridge University Press, \doi{10.1017/9781009157896.009}, 2021.

\bibitem[{Fuglestvedt and Haerter(2020)}]{fuglestvedt_coldpools_2020}
Fuglestvedt, H.~F. and Haerter, J.~O.: Cold Pools as Conveyor Belts of Moisture, Geophysical Research Letters, 47, e2020GL087\,319, \doi{10.1029/2020GL087319}, 2020.

\bibitem[{Grant et~al.(2022)Grant, van~den Heever, Haddad, Bukowski, Marinescu, Storer, Posselt, and Stephens}]{grant_linear_2022}
Grant, L.~D., van~den Heever, S.~C., Haddad, Z.~S., Bukowski, J., Marinescu, P.~J., Storer, R.~L., Posselt, D.~J., and Stephens, G.~L.: A Linear Relationship between Vertical Velocity and Condensation Processes in Deep Convection, Journal of the Atmospheric Sciences, 79, 449--466, \doi{10.1175/JAS-D-21-0035.1}, 2022.

\bibitem[{Grundner et~al.(2024)Grundner, Beucler, Gentine, and Eyring}]{grundner_data-driven_2024}
Grundner, A., Beucler, T., Gentine, P., and Eyring, V.: Data-Driven Equation Discovery of a Cloud Cover Parameterization, Journal of Advances in Modeling Earth Systems, 16, e2023MS003\,763, \doi{10.1029/2023MS003763}, 2024.

\bibitem[{H{\"a}fner et~al.(2023)H{\"a}fner, Gemmrich, and Jochum}]{hafner_rogue_2023}
H{\"a}fner, D., Gemmrich, J., and Jochum, M.: Machine-guided discovery of a real-world rogue wave model, Proceedings of the National Academy of Sciences, 120, e2306275\,120, \doi{10.1073/pnas.2306275120}, 2023.

\bibitem[{Hannah(2017)}]{hannah2017entrainment}
Hannah, W.~M.: Entrainment versus dilution in tropical deep convection, Journal of the Atmospheric Sciences, 74, 3725--3747, 2017.

\bibitem[{Held et~al.(1993)Held, Hemler, and Ramaswamy}]{held_rce_1993}
Held, I.~M., Hemler, R.~S., and Ramaswamy, V.: Radiative-Convective Equilibrium with Explicit Two-Dimensional Moist Convection, Journal of Atmospheric Sciences, 50, 3909 -- 3927, \doi{10.1175/1520-0469(1993)050<3909:RCEWET>2.0.CO;2}, 1993.

\bibitem[{Hernández-Deckers and Sherwood(2016)}]{hernandez-deckers_numerical_2016}
Hernández-Deckers, D. and Sherwood, S.~C.: A Numerical Investigation of Cumulus Thermals, Journal of the Atmospheric Sciences, 73, 4117--4136, \doi{10.1175/JAS-D-15-0385.1}, 2016.

\bibitem[{Hogan et~al.(2006)Hogan, Mittermaier, and Illingworth}]{hogan_icewater_2008}
Hogan, R.~J., Mittermaier, M.~P., and Illingworth, A.~J.: The Retrieval of Ice Water Content from Radar Reflectivity Factor and Temperature and Its Use in Evaluating a Mesoscale Model, Journal of Applied Meteorology and Climatology, 45, 301--317, \doi{10.1175/JAM2340.1}, 2006.

\bibitem[{Hohenegger et~al.(2020)Hohenegger, Kornblueh, Klocke, Becker, Cioni, Engels, Schulzweida, and Stevens}]{hohenegger2020climate}
Hohenegger, C., Kornblueh, L., Klocke, D., Becker, T., Cioni, G., Engels, J.~F., Schulzweida, U., and Stevens, B.: Climate statistics in global simulations of the atmosphere, from 80 to 2.5 km grid spacing, Journal of the Meteorological Society of Japan. Ser. II, 98, 73--91, 2020.

\bibitem[{Jakob et~al.(2019)Jakob, Singh, and Jungandreas}]{jakob_radiative_2019}
Jakob, C., Singh, M., and Jungandreas, L.: Radiative convective equilibrium and organized convection: An observational perspective, Journal of Geophysical Research: Atmospheres, 124, 5418--5430, 2019.

\bibitem[{Jeevanjee(2017)}]{jeevanjee2017vertical}
Jeevanjee, N.: Vertical velocity in the gray zone, Journal of Advances in Modeling Earth Systems, 9, 2304--2316, 2017.

\bibitem[{Jeevanjee and Romps(2016)}]{jeevanjee_effective_2016}
Jeevanjee, N. and Romps, D.~M.: Effective buoyancy at the surface and aloft, Quarterly Journal of the Royal Meteorological Society, 142, 811--820, \doi{10.1002/qj.2683}, 2016.

\bibitem[{Jeevanjee et~al.(2022)Jeevanjee, Held, and Ramaswamy}]{jeevanjee2022manabe}
Jeevanjee, N., Held, I., and Ramaswamy, V.: Manabe's Radiative--Convective Equilibrium, Bulletin of the American Meteorological Society, 103, E2559--E2569, \doi{10.1175/BAMS-D-21-0351.1}, 2022.

\bibitem[{Johns and Doswell~III(1992)}]{johns_severe_1992}
Johns, R.~H. and Doswell~III, C.~A.: Severe local storms forecasting, Weather and Forecasting, 7, 588--612, 1992.

\bibitem[{Kaptanoglu et~al.(2022)Kaptanoglu, de~Silva, Fasel, Kaheman, Goldschmidt, Callaham, Delahunt, Nicolaou, Champion, Loiseau, Kutz, and Brunton}]{kaptanoglu_pysindy_2022}
Kaptanoglu, A.~A., de~Silva, B.~M., Fasel, U., Kaheman, K., Goldschmidt, A.~J., Callaham, J., Delahunt, C.~B., Nicolaou, Z.~G., Champion, K., Loiseau, J.-C., Kutz, J.~N., and Brunton, S.~L.: PySINDy: A comprehensive Python package for robust sparse system identification, Journal of Open Source Software, 7, 3994, \doi{10.21105/joss.03994}, 2022.

\bibitem[{Khairoutdinov and Randall(2003)}]{khairoutdinov_cloud_2003}
Khairoutdinov, M.~F. and Randall, D.~A.: Cloud Resolving Modeling of the ARM Summer 1997 IOP: Model Formulation, Results, Uncertainties, and Sensitivities, Journal of the Atmospheric Sciences, 60, 607--625, \doi{10.1175/1520-0469(2003)060<0607:CRMOTA>2.0.CO;2}, 2003.

\bibitem[{Kingma and Ba(2014)}]{kingma_adam_2014}
Kingma, D.~P. and Ba, J.: Adam: A method for stochastic optimization, arXiv preprint arXiv:1412.6980, 2014.

\bibitem[{Kuo and Neelin(2025)}]{kuo_anelastic_2025}
Kuo, Y.-H. and Neelin, J.~D.: Anelastic Convective Entities. Part {II}: Adjustment Processes and Convective Cold Top, Journal of the Atmospheric Sciences, \doi{10.1175/JAS-D-24-0130.1}, 2025.

\bibitem[{Leger et~al.(2019)Leger, Lafore, Piriou, and Guérémy}]{leger_simple_2019}
Leger, J., Lafore, J.-P., Piriou, J.-M., and Guérémy, J.-F.: A Simple Model of Convective Drafts Accounting for the Perturbation Pressure Term, Journal of the Atmospheric Sciences, 76, 3129--3149, \doi{10.1175/JAS-D-18-0281.1}, 2019.

\bibitem[{Manabe and Strickler(1964)}]{manabe_thermal_1964}
Manabe, S. and Strickler, R.~F.: Thermal equilibrium of the atmosphere with a convective adjustment, J. Atmos. Sci., 21, 361--385, publisher: American Meteorological Society, 1964.

\bibitem[{Morrison(2016)}]{Morrison2016}
Morrison, H.: Impacts of Updraft Size and Dimensionality on the Perturbation Pressure and Vertical Velocity in Cumulus Convection. Part II: Comparison of Theoretical and Numerical Solutions and Fully Dynamical Simulations, Journal of the Atmospheric Sciences, 73, 1455--1480, \doi{10.1175/JAS-D-15-0041.1}, 2016.

\bibitem[{Morrison et~al.(2020)Morrison, Peters, Varble, Hannah, and Giangrande}]{morrison2020thermal}
Morrison, H., Peters, J.~M., Varble, A.~C., Hannah, W.~M., and Giangrande, S.~E.: Thermal chains and entrainment in cumulus updrafts. Part I: Theoretical description, Journal of the Atmospheric Sciences, 77, 3637--3660, 2020.

\bibitem[{Morrison et~al.(2022)Morrison, Jeevanjee, and Yano}]{morrison2022dynamic}
Morrison, H., Jeevanjee, N., and Yano, J.-I.: Dynamic pressure drag on rising buoyant thermals in a neutrally stable environment, Journal of the Atmospheric Sciences, 79, 3045--3063, 2022.

\bibitem[{Moser and Lasher-Trapp(2017)}]{moser2017influence}
Moser, D.~H. and Lasher-Trapp, S.: The influence of successive thermals on entrainment and dilution in a simulated cumulus congestus, Journal of the Atmospheric Sciences, 74, 375--392, 2017.

\bibitem[{Muller and Takayabu(2020)}]{muller_extremes_2020}
Muller, C. and Takayabu, Y.: Response of precipitation extremes to warming: what have we learned from theory and idealized cloud-resolving simulations, and what remains to be learned?, Environmental Research Letters, 15, 035\,001, \doi{10.1088/1748-9326/ab7130}, 2020.

\bibitem[{Muller and O’Gorman(2011)}]{muller2011energetic}
Muller, C.~J. and O’Gorman, P.: An energetic perspective on the regional response of precipitation to climate change, Nature Climate Change, 1, 266--271, 2011.

\bibitem[{O'Gorman and Schneider(2009)}]{ogorman_schneider_2009}
O'Gorman, P.~A. and Schneider, T.: The physical basis for increases in precipitation extremes in simulations of 21st-century climate change, Proceedings of the National Academy of Sciences, 106, 14\,773--14\,777, \doi{10.1073/pnas.0907610106}, 2009.

\bibitem[{Parodi and Emanuel(2009)}]{parodi_theory_2009}
Parodi, A. and Emanuel, K.: A Theory for Buoyancy and Velocity Scales in Deep Moist Convection, Journal of the Atmospheric Sciences, 66, 3449--3463, \doi{10.1175/2009JAS3103.1}, 2009.

\bibitem[{Pechacova et~al.(2026)Pechacova, Casallas, Beucler, Agasthya, and Muller}]{pechacova_2026_zenodo}
Pechacova, B., Casallas, A., Beucler, T., Agasthya, L., and Muller, C.: Code and data for: Maximum updraft velocity beyond CAPE: the role of boundary layer dynamics and pressure perturbation, \doi{10.5281/zenodo.21474646}, 2026.

\bibitem[{Peters et~al.(2020)Peters, Morrison, Varble, Hannah, and Giangrande}]{peters2020thermal}
Peters, J.~M., Morrison, H., Varble, A.~C., Hannah, W.~M., and Giangrande, S.~E.: Thermal chains and entrainment in cumulus updrafts. Part II: Analysis of idealized simulations, Journal of the Atmospheric Sciences, 77, 3661--3681, 2020.

\bibitem[{Peters et~al.(2023)Peters, Chavas, Su, Morrison, and Coffer}]{peters_analytic_2023}
Peters, J.~M., Chavas, D.~R., Su, C.-Y., Morrison, H., and Coffer, B.~E.: An Analytic Formula for Entraining CAPE in Midlatitude Storm Environments, Journal of the Atmospheric Sciences, 80, 2165 -- 2186, \doi{10.1175/JAS-D-23-0003.1}, 2023.

\bibitem[{Robinson et~al.(2025)Robinson, Kirshbaum, Sherwood, Cahill, Juliano, and Liu}]{Robinson2025}
Robinson, F., Kirshbaum, D.~J., Sherwood, S.~C., Cahill, L., Juliano, E., and Liu, C.: Investigating the Effects of Orography and Ambient Wind on Deep Convection over Tropical Islands, Journal of the Atmospheric Sciences, 82, 139--157, \doi{10.1175/JAS-D-23-0204.1}, 2025.

\bibitem[{Romps(2019)}]{romps_evaluating_2019}
Romps, D.~M.: Evaluating the future of lightning in cloud-resolving models, Geophysical Research Letters, 46, 14\,863--14\,871, \doi{10.1029/2019GL085748}, 2019.

\bibitem[{Romps and Charn(2015)}]{romps_sticky_2015}
Romps, D.~M. and Charn, A.~B.: Sticky Thermals: Evidence for a Dominant Balance between Buoyancy and Drag in Cloud Updrafts, Journal of the Atmospheric Sciences, 72, 2890--2901, \doi{10.1175/JAS-D-15-0042.1}, 2015.

\bibitem[{Segura et~al.(2025)Segura, Pedruzo-Bagazgoitia, Weiss, M\"uller, Rackow, Lee, Dolores-Tesillos, Benedict, Aengenheyster, Aguridan, Arduini, Baker, Bao, Bastin, Baulenas, Becker, Beyer, Bockelmann, Br\"uggemann, Brunner, Cheedela, Das, Denissen, Dragaud, Dziekan, Ekblom, Engels, Esch, Forbes, Frauen, Freischem, Garc\'{\i}a-Maroto, Geier, Gierz, Gonz\'alez-Cervera, Grayson, Griffith, Gutjahr, Haak, Hadade, Haslehner, ul~Hasson, Hegewald, Kluft, Koldunov, Koldunov, K\"olling, Koseki, Kosukhin, Kousal, Kuma, Kumar, Li, Maury, Meindl, Milinski, Mogensen, Niraula, Nowak, Praturi, Proske, Putrasahan, Redler, Santuy, S\'arm\'any, Schnur, Scholz, Sidorenko, Sp\"at, S\"utzl, Takasuka, Tompkins, Uribe, Valentini, Veerman, Voigt, Warnau, Wachsmann, Wac{\l}awczyk, Wedi, Wieners, Wille, Winkler, Wu, Ziemen, Zimmermann, Bender, Bojovic, Bony, Bordoni, Brehmer, Dengler, Dutra, Faye, Fischer, van Heerwaarden, Hohenegger, J\"arvinen, Jochum, Jung, Jungclaus, Keenlyside, Klocke, Konow, Klose, Malinowski, Martius,
  Mauritsen, Mellado, Mieslinger, Mohino, Paw{\l}owska, Peters-von Gehlen, Sarr\'e, Sobhani, Stier, Tuppi, Vidale, Sandu, and Stevens}]{segura_2025_nextgems}
Segura, H., Pedruzo-Bagazgoitia, X., Weiss, P., M\"uller, S.~K., Rackow, T., Lee, J., Dolores-Tesillos, E., Benedict, I., Aengenheyster, M., Aguridan, R., Arduini, G., Baker, A.~J., Bao, J., Bastin, S., Baulenas, E., Becker, T., Beyer, S., Bockelmann, H., Br\"uggemann, N., Brunner, L., Cheedela, S.~K., Das, S., Denissen, J., Dragaud, I., Dziekan, P., Ekblom, M., Engels, J.~F., Esch, M., Forbes, R., Frauen, C., Freischem, L., Garc\'{\i}a-Maroto, D., Geier, P., Gierz, P., Gonz\'alez-Cervera, A., Grayson, K., Griffith, M., Gutjahr, O., Haak, H., Hadade, I., Haslehner, K., ul~Hasson, S., Hegewald, J., Kluft, L., Koldunov, A., Koldunov, N., K\"olling, T., Koseki, S., Kosukhin, S., Kousal, J., Kuma, P., Kumar, A.~U., Li, R., Maury, N., Meindl, M., Milinski, S., Mogensen, K., Niraula, B., Nowak, J., Praturi, D.~S., Proske, U., Putrasahan, D., Redler, R., Santuy, D., S\'arm\'any, D., Schnur, R., Scholz, P., Sidorenko, D., Sp\"at, D., S\"utzl, B., Takasuka, D., Tompkins, A., Uribe, A., Valentini, M., Veerman, M.,
  Voigt, A., Warnau, S., Wachsmann, F., Wac{\l}awczyk, M., Wedi, N., Wieners, K.-H., Wille, J., Winkler, M., Wu, Y., Ziemen, F., Zimmermann, J., Bender, F. A.-M., Bojovic, D., Bony, S., Bordoni, S., Brehmer, P., Dengler, M., Dutra, E., Faye, S., Fischer, E., van Heerwaarden, C., Hohenegger, C., J\"arvinen, H., Jochum, M., Jung, T., Jungclaus, J.~H., Keenlyside, N.~S., Klocke, D., Konow, H., Klose, M., Malinowski, S., Martius, O., Mauritsen, T., Mellado, J.~P., Mieslinger, T., Mohino, E., Paw{\l}owska, H., Peters-von Gehlen, K., Sarr\'e, A., Sobhani, P., Stier, P., Tuppi, L., Vidale, P.~L., Sandu, I., and Stevens, B.: nextGEMS: entering the era of kilometer-scale Earth system modeling, Geoscientific Model Development, 18, 7735--7761, \doi{10.5194/gmd-18-7735-2025}, 2025.

\bibitem[{Shamekh et~al.(2026)Shamekh, Angulo-Umana, and O’Gorman}]{shamekh_rainarea_2026}
Shamekh, S., Angulo-Umana, P., and O’Gorman, P.~A.: Data-Driven Modeling of Stratiform and Convective Rain Area, Journal of the Atmospheric Sciences, 83, 961 -- 980, \doi{10.1175/JAS-D-25-0178.1}, 2026.

\bibitem[{Sherwood et~al.(2013)Sherwood, Hernández-Deckers, Colin, and Robinson}]{sherwood_slippery_2013}
Sherwood, S.~C., Hernández-Deckers, D., Colin, M., and Robinson, F.: Slippery Thermals and the Cumulus Entrainment Paradox*, Journal of the Atmospheric Sciences, 70, 2426--2442, \doi{10.1175/JAS-D-12-0220.1}, 2013.

\bibitem[{Singh and O'Gorman(2013)}]{singh_influence_2013}
Singh, M.~S. and O'Gorman, P.~A.: Influence of entrainment on the thermal stratification in simulations of radiative-convective equilibrium, Geophysical Research Letters, 40, 4398--4403, \doi{10.1002/grl.50796}, 2013.

\bibitem[{Singh and O'Gorman(2015)}]{singh_increases_2015}
Singh, M.~S. and O'Gorman, P.~A.: Increases in moist-convective updraught velocities with warming in radiative-convective equilibrium, Quarterly Journal of the Royal Meteorological Society, 141, 2828--2838, \doi{10.1002/qj.2567}, 2015.

\bibitem[{Stephens et~al.(2008)Stephens, Vane, Tanelli, Im, Durden, Rokey, Reinke, Partain, Mace, Austin, L'Ecuyer, Haynes, Lebsock, Suzuki, Waliser, Wu, Kay, Gettelman, Wang, and Marchand}]{stephens_cloudsat_2008}
Stephens, G.~L., Vane, D.~G., Tanelli, S., Im, E., Durden, S., Rokey, M., Reinke, D., Partain, P., Mace, G.~G., Austin, R., L'Ecuyer, T., Haynes, J., Lebsock, M., Suzuki, K., Waliser, D., Wu, D., Kay, J., Gettelman, A., Wang, Z., and Marchand, R.: CloudSat mission: Performance and early science after the first year of operation, Journal of Geophysical Research, 113, D00A18, \doi{10.1029/2008JD009982}, 2008.

\bibitem[{Tompkins(2001)}]{tompkins_coldpools_2001}
Tompkins, A.: {Organization of Tropical Convection in Low Vertical Wind Shears: The Role of Cold Pools.}, Journal of Atmospheric Sciences, 58, 1650--1672, \doi{10.1175/1520-0469(2001)058<1650:OOTCIL>2.0.CO;2}, 2001.

\bibitem[{Torri et~al.(2015)Torri, Kuang, and Tian}]{torri_mechanisms_2015}
Torri, G., Kuang, Z., and Tian, Y.: Mechanisms for convection triggering by cold pools, Geophysical Research Letters, 42, 1943--1950, \doi{10.1002/2015GL063227}, 2015.

\bibitem[{van~der Walt et~al.(2014)van~der Walt, {S}ch\"onberger, {Nunez-Iglesias}, {B}oulogne, {W}arner, {Y}ager, {G}ouillart, {Y}u, and the scikit-image contributors}]{scikit-image}
van~der Walt, S., {S}ch\"onberger, J.~L., {Nunez-Iglesias}, J., {B}oulogne, F., {W}arner, J.~D., {Y}ager, N., {G}ouillart, E., {Y}u, T., and the scikit-image contributors: scikit-image: image processing in {P}ython, PeerJ, 2, e453, \doi{10.7717/peerj.453}, 2014.

\bibitem[{Wing et~al.(2018)Wing, Reed, Satoh, Stevens, Bony, and Ohno}]{wing_rcemip_2018}
Wing, A.~A., Reed, K.~A., Satoh, M., Stevens, B., Bony, S., and Ohno, T.: Radiative--convective equilibrium model intercomparison project, Geoscientific Model Development, 11, 793--813, 2018.

\bibitem[{Zanna and Bolton(2020)}]{zanna_ocean_2020}
Zanna, L. and Bolton, T.: Data-Driven Equation Discovery of Ocean Mesoscale Closures, Geophysical Research Letters, 47, e2020GL088\,376, \doi{10.1029/2020GL088376}, 2020.

\bibitem[{Zhang(2009)}]{zhang_effects_2009}
Zhang, G.~J.: Effects of entrainment on convective available potential energy and closure assumptions in convection parameterization, Journal of Geophysical Research: Atmospheres, 114, \doi{10.1029/2008JD010976}, 2009.

\bibitem[{Zhou and Xie(2019)}]{zhou_spectralplume_2019}
Zhou, W. and Xie, S.-P.: A Conceptual Spectral Plume Model for Understanding Tropical Temperature Profile and Convective Updraft Velocities, Journal of the Atmospheric Sciences, 76, 2801--2814, \doi{10.1175/JAS-D-18-0330.1}, 2019.

\bibitem[{Zhu et~al.(1997)Zhu, Byrd, Lu, and Nocedal}]{zhu_lbfgsb_1997}
Zhu, C., Byrd, R.~H., Lu, P., and Nocedal, J.: Algorithm 778: {L-BFGS-B}, ACM Trans. Math. Softw., 23, 550--560, \doi{10.1145/279232.279236}, 1997.

\bibitem[{Zipser and LeMone(1980)}]{zipser_cumulonimbus_1980}
Zipser, E.~J. and LeMone, M.~A.: Cumulonimbus Vertical Velocity Events in GATE. Part II: Synthesis and Model Core Structure, Journal of Atmospheric Sciences, 37, 2458 -- 2469, \doi{10.1175/1520-0469(1980)037<2458:CVVEIG>2.0.CO;2}, 1980.

\end{thebibliography}

\end{document}